\documentclass[14pt,manuscript]{aastex}

\shorttitle{DPOSS Compact Groups}
\shortauthors{de Carvalho et al.}

\begin{document}

\title{A Catalog of Distant Compact Groups Using DPOSS}

\author{R.\ R. de Carvalho}
\affil{INPE/MCT, Av. dos Astronautas 1758, S.J.Campos - SP 12227-010 Brazil}
\email{reinaldo@das.inpe.br}

\author{T.\ S. Gon\c calves}
\affil{Palomar Observatory, Caltech, MS105-24, Pasadena, CA 91125}
\email{tsg@astro.caltech.edu}

\author{A. Iovino}
\affil{Osservatorio Astronomico di Brera, via Brera 28, I-20121 Milano, Italy}
\email{iovino@brera.mi.astro.it}

\author{J.\ L. Kohl-Moreira}
\affil{Observat\'orio Nacional, Rua General Jos\'e Cristino, 77, 20921-400, S\~ao Crist\'ov\~ao,
Rio de Janeiro, Brasil}
\email{kohl@on.br}

\author{R.\ R. Gal} \affil{Dept. of Physics, UC - Davis, One Shields Ave., Davis, CA 95616}
\email{gal@physics.ucdavis.edu}

\author{S.\ G. Djorgovski}
\affil{Palomar Observatory, Caltech, MS105-24, Pasadena, CA 91125}
\email{george@astro.caltech.edu} 

\begin{abstract}

In this paper we present an objectively defined
catalog of 459 small, high density groups of galaxies out to $ z\sim
0.2$ in a region of $\sim 6260$ square degrees in the northern sky
derived from the Digitized Second Palomar Observatory Sky Survey. Our
catalog extends down to $r = 19.0$ and has a median redshift of
$z_{\rm med}$ = 0.12, making it complementary to Hickson's catalog for
the nearby universe ($z_{\rm med}$ = 0.03). The depth and angular
coverage of this catalog makes it valuable for studies of the general
characteristics of small groups of galaxies and how galaxies evolve in
and around them.  We also examine the relationship between compact
groups and large scale structure.
  
\end{abstract}

\keywords{galaxies: clusters of --- galaxies: surveys --- methods:
statistical}

\section{Introduction}

Compact groups of galaxies (CGs), as the name suggests, are systems
composed of a small number of galaxies in an angularly compact
configuration on the sky. Stephan's quintet is the earliest known
member of this class, discovered in 1877 by Edouard Stephan. In 1948,
the Seyfert sextet was found to belong to the same family. Through the
nineteen-seventies, only visual detections of these systems were
possible and no homogeneous sample was available for systematic
studies of these low-mass structures. For a brief history of the
heterogeneous samples of small groups see Iovino et al. (2003).

Rose (1977) was the first to attempt construction of an objective
sample of CGs in the northern sky. By imposing a minimum requirement
of three galaxies brighter than magnitude 17.5 in B-band with a
density contrast of at least 1000, Rose found 170 triplets, 33
quartets, and 2 quintets over an area of $\sim$3100 degrees in the
northern hemisphere. However, the most studied sample of CGs is that
defined by Hickson (1982).  Instead of using a fixed limiting
magnitude, Hickson required that four or more galaxies have a maximum
magnitude difference of 3.0 mag.  The group's surface brightness was
defined as $\mu_{\rm gr} = -2.5~{\rm log}((\Sigma~ 10^{-0.4m_{\rm
i}})/ \pi \rm R_{\rm gr}^{2})$, where the $m_{\rm i}$'s are the
magnitudes of the galaxies defining the group. He required a limiting
surface brightness of 26.0 mag arcsec$^{-2}$ in the E-band (similar to
Johnson R). He also imposed an encircling null ring where no galaxies
are present of area equal to 8$\pi{\rm R_{\rm G}}^{2}$, where ${\rm
R}_{\rm G}$ is the radius of the smallest circle encompassing the
centers of the group members. With these constraints, Hickson found
100 CGs, covering $\sim$27000 square degrees on the sky, mostly in the
north with a small extension to the south. Many studies have been
performed using the Hickson sample (see Hickson 1997 for a
review). From the spectroscopic survey presented in Hickson et
al. (1992), out of the 100 CGs, only 69 were found to have four or
more concordant galaxies, while 23 were triplets, with a median
redshift $z_{\rm med}$ of 0.03 for these 92 galactic systems.

Several studies were conducted with the goal of understanding the
physical nature of the CGs defined in the Hickson sample, which,
despite being affected by incompleteness and bias (Prandoni et
al. 1994), is comprised of interesting systems. An important question
is whether or not these systems represent real dynamical structures
(Hickson 1997).  A critical point is that every sample selected based
on an apparent density contrast will suffer from projection effects,
such as a tendency to select prolate groups radially oriented along
the line of sight (Plionis, Basilakos, \& Tovmassian 2004), or even
transient configurations (Rose 1979).

N-body simulations show that CGs might form a single giant elliptical
over a few crossing times through mergers (e.g.  Barnes 1985, 1989;
Mamon 1987; Zheng, Valtonen, \& Chernin 1993). However, CGs are
observed in the local universe, which must be reconciled with these
simulations.  Diaferio et al. (1994) addressed this issue through a
set of N-body simulations where the initial conditions emulated the
properties of sparse groups. They find that although the CG lifetime
seems to be short ($\sim$ 1 Gyr), the system can be constantly
replenished by neighboring galaxies. These simulations succeeded in
recovering observed properties of CGs including the morphological
distribution with respect to the field population, the dynamical
properties, and above all showed that most of the group members are
not the result of mergers. A criticism of this scenario
(e.g. Athanassoula 2000) is the absence of possible fossils in the
field population, since in the Diaferio et al. (1994) model after a
considerable fraction of the Hubble time the more massive sparse
groups become an isolated giant elliptical. This problem could be
alleviated by the discovery of the so-called fossil groups (Ponman et
al. 1994 and Jones et al. 2003) left behind after a group
merges: an elliptical embedded in an overluminous X-ray halo. A
significant fraction of the group mass may also reside in a common
halo outside the galaxies (see Zabludoff and Mulchaey 1998),
therefore significantly slowing the rate at which the group members
interact and eventually merge (see Governato et al. 1991).

The question is then how we distinguish between an elliptical
originating in the field and one resulting from the merger of a
sparse group. Recently, de la Rosa, de Carvalho, \& Zepf (2001a), by
studying the fundamental plane of elliptical galaxies in the field and
in CGs, concluded that both families are similar from the structural
and dynamical points of view. Proctor et al. (2004) and de la Rosa et
al. (2001b), on the other hand, found that ellipticals in CGs are more
similar to those in clusters than the ones in the field, based on
analysis of their stellar population properties.  de Carvalho et
al. (1997) and Ribeiro et al. (1998), studying the vicinity of 17 CGs
from the Hickson sample, out to $\sim$9000 km s$^{-1}$, found that CGs
can be divided into three categories: 1) sparse groups; 2) systems
with a core and a halo; and 3) real compact groups. They find that
each category has its own characteristic surface density profile,
which reinforces the view that CGs compose a very heterogeneous family
and not a single type of system. This result not only supports the
scenario envisaged by Diaferio et al. (1994) but also prompted us to
investigate how CGs are distributed relative to large scale structure,
including loose groups and poor and rich clusters.

Several studies examined the relation between CGs and other
structures. Rood \& Struble (1994) looked for structures within 1.0
Mpc (transverse) of the CGs, with a difference in radial velocity at
most four times the group's velocity dispersion. They find that 75\%
of Hickson's groups are near other structures such as loose groups and
Abell clusters, which might indicate that CGs are effectively part of
the same hierarchy we observe in the universe, from isolated galaxies
to superclusters.  Similar work was done by Kelm \& Focardi (2004)
based on the Updated Zwicky Catalog. They generate a catalog of CGs
which is quite distinct from that of Hickson, using redshifts to
define galaxy aggregates. They studied the morphological
characteristics of the galaxies in the vicinity of the groups using a
region between 0.2 $h^{-1}$ Mpc and 1.0 $h^{-1}$ Mpc from the group
center, and requiring a velocity difference less than 1000 km
s$^{-1}$. These neighbor galaxies exhibit an intermediate nature
between field and group galaxies, reinforcing the finding that CGs are
a distinct physical entity and not merely the result of projection
effects. Moreover, excluding the brightest group galaxies from the
comparison, the difference in morphology distribution disappears,
indicating that the group galaxy population is more evolved than field
galaxies.  Helsdon \& Ponman (2000a,b) offer a different perspective
by combining CGs and loose groups into a single sample and examining
their X-ray properties.  They suggest that both clustering scales
should be considered together since their X-ray properties are very
similar.

Although the importance of CGs in linking AGN and environment may
appear obvious, only recently has the level of nuclear activity (AGN)
and starburst activity among compact group galaxies been
systematically measured (Coziol et al. 1998a, 1998b). Based on an
analysis of 82 bright galaxies from 17 CGs, these authors found that
AGN are mainly located in early-type and luminous systems, which are
similar to their hosts in the field (Phillips et al. 1986). Coziol et
al. (1998a) also found that AGN are more concentrated towards the
central parts of the groups, suggesting a morphology-density-activity
relation. This finding was later confirmed by Coziol, Iovino, \& de
Carvalho (2000) studying a sample of CGs defined by Iovino (2002). This
may prove to be an important diagnostic of how dynamically evolved a
CG is. Several other works examined the question of how the
environment is related to the physical mechanism responsible for the
feeding of galactic nuclei (Shimada et al.  2000, Schmitt
2001, Coziol et al. 2004, Kelm, Focardi, \& Zitelli 2004).

Recently, two large samples of CGs were published. The first was
selected from the Digitized Second Palomar Observatory Sky Survey
(DPOSS) and based on plate data (Iovino et al. 2003), while the second
by Lee et al. (2004) used the Sloan Digital Sky Survey (SDSS) and was
based on CCD data. When comparing both samples it is important to
consider the methodology employed to define a small group of galaxies,
which is always a difficult task.  The main advantage of both
contributions over all previous works is the objective nature of the
algorithms used to select the systems, making comparisons with models
and between the surveys more meaningful. A similar approach was used
for the first time in Prandoni et al. (1994), and Iovino (2002) also
defines a sample of CGs in a totally automated fashion.

Compact groups are thought to be ideal places to study galaxy
evolution, since their high density and low velocity dispersion
naively imply a higher merging rate. The observational evidence from
nearby CGs shows that this is not necessarily true in a majority of
cases. Thus, better understanding of the weaknesses of both local and
higher redshift samples is necessary to distinguish reality from these
naive assumptions. In this paper, we present an extension of the
sample from Iovino et al. (2003). Our final list of 459 CGs covers
6260 square degrees of the northern sky, with galactic latitude
restricted to $|b| > 40^{\circ}$, with a small extension to the south, and
extends to $z\sim$0.2. In Section 2 we present the data and method
used to construct the sample. In Section 3 we discuss the global
properties of the sample and possible biases, while in Section 4 some
basic CG properties are compared to those of the field galaxy
population. Section 5 presents a preliminary comparison between the
CGs and the large scale distribution of other structures. Finally, in
Section 6, we summarize our main findings. The most favored cosmology
today is used throughout the paper, $\Omega_{\rm m}$ = 0.3,
$\Omega_{\lambda}$ = 0.7, and H$_{\circ}$ = 67 km s$^{-1}$ Mpc$^{-1}$.
 
\section{Data and Algorithm Used}

\subsection{The Galaxy Catalog}

The catalog of CG candidates presented in this work was constructed
using the galaxy catalogs from DPOSS. The photographic plates from
POSS-II cover the whole northern sky, including a small region of the
southern sky ($\delta > -3^{\circ}$), with 894 fields of
6.5$^{\circ}\times6.5^{\circ}$, and 5$^{\circ}$ separation between
plate centers. This provides a large area of intersection between
plates allowing strict control of the instrumental magnitude
system. Plates were taken in three photometric bands: J
(IIIa-J$+$GG395, $\lambda_{\rm eff} \sim 480 nm$), F (IIIa-F$+$RG610,
$\lambda_{\rm eff} \sim 650 nm$), and N (IV-N$+$RG9, $\lambda_{\rm
eff} \sim 850 nm$). The limiting plate magnitudes are B$_{\rm J}\sim
22.5^{\rm m}$, R$_{\rm F}\sim 20.8^{\rm m}$ and I$_{\rm N}\sim
19.5^{\rm m}$, respectively. These limits correspond, in the
Thuan-Gunn $gri$ system, to 21.5$^{\rm m}$, 20.5$^{\rm m}$, and
19.8$^{\rm m}$, respectively. Digitization was performed by a
microdensitometer PDS modified to yield high photometric quality. Each
pixel has 15$\times$15$\mu{\rm m}$, corresponding to 1$\times$1
arcsec. Early processing of DPOSS is described by Weir et
al. (1995a,b,c).  An overview of the DPOSS survey is given in
Djorgovski et al. (1999), and photometric calibration is described in
detail in Gal et al. (2004).

An important step in defining a catalog of detected objects is
star-galaxy separation. For this, we have used two independent methods
to perform automated image classification: a decision tree and an
artificial neural network, described in detail in Fayyad (1991) and
Odewahn et al. (1992), respectively. The two algorithms make use of
the same set of attributes and their final accuracies are comparable
(see Odewahn et al. 2004), with 10\% contamination at $r = 19.5$
mag. Further information on the star-galaxy separation methods can be
found in Odewahn et al. (2004).

\subsection{The Method}

The first sample of CGs based on DPOSS data was the Palomar Compact Group
catalog (PCG, Iovino et al. 2003), which defined CGs over an area of
2000 square degrees around the north Galactic pole using galaxies
between $r=16$ mag and $r=20$ mag. The brightest galaxies in the
groups had magnitudes ranging between 16.0 and 17.0 mag.
The method used in this work is conceptually the
same as the one employed by Iovino et al. (2003) and we refer the
reader to that paper for more specific details of the algorithm.

We summarize here the most important points of the selection criteria
used in the search for small aggregates of galaxies:

\begin{itemize} 

\item{ --- richness: $n_{\rm memb} \; \ge \; 4$ in the magnitude
       interval $\Delta\rm mag_{comp} = m_{faintest}-m_{brightest}$,
       with the constraint $\Delta\rm mag_{comp} \leq 2^m $. This
       magnitude difference is considerably stricter than Hickson's
       (3.0$^{\rm m}$), thus maintaining a low contamination rate but 
       reducing completeness.}

\item{ --- isolation: $\rm R_{isol} \; \ge \; 3R_{gr}$, where $\rm
      R_{isol}$ is the distance from the center of the smallest circle
      encompassing all of a group's galaxies to the nearest
      non--member galaxy within 0.5 magnitudes of the faintest group
      member. This criterion avoids finding small aggregates within a
      larger structure, such as a cluster.}

\item{ --- compactness: $\mu_{\rm gr} \; < \; \mu_{\rm limit}$, where
      $\mu_{\rm gr}$ is the mean surface brightness within the circle
      of radius $\rm R_{gr}$, and $\mu_{\rm limit}$ = 24.0 mag
      arcsec$^{-2}$ in $r$-band.}
\end{itemize} 

\noindent Here $n_{\rm memb}$ is the number of member galaxies.
For comparison, Hickson (1982) used $\mu_{\rm limit}$ = 26 mag arcsec$^{-2}$ in the
red E band of POSS-I.

Through the use of simulations, Iovino et al. (2003) have shown that a
10\% contamination rate is expected. It is interesting to note that
from the initially selected 100 groups by Hickson (1982), 92 had more
than 3 concordant redshifts, and from these 92 only 69 had at least 4
members. Thus, triplets comprise $\sim$25\% of his sample, a
considerable contamination rate. Although the surface brightness
constraint in our catalog is much more restrictive than the one used
by Hickson, there may still be a sizable amount of contamination by
triplets. The contamination rate can only be measured by carrying out
a complete redshift survey, as was the case for the Hickson catalog.

\section{A New Sample For the Northern Hemisphere and its General Properties}

The sample described in this paper was obtained by applying the same
algorithm used in Iovino et al. (2003). However, a number of
complications required special attention. Star forming regions in
spiral galaxies sometimes appear in the galaxy catalog as multiple
isolated objects, and are detected as a CG. These obviously spurious
objects were removed by visual inspection.  In addition, the overlap
regions of adjacent plates result in duplicate detections of some CGs,
which were also automatically removed. In these cases the CG in the
final catalog is the detection with the larger number of members.
Finally, in some cases different sets of galaxies over a small region
satisfy the criteria for forming a group, resulting in two detections
of the same group with different galaxies. We again discarded the
group with fewer members, or in those cases where the number of
members was the same we eliminated the instance with smaller $\Delta
\rm mag_{\rm comp}$. None of the above scenarios affect the
objectiveness of the final sample, but demonstrate that careful
examination of the sample is of paramount importance in understanding
the selection function and avoiding obvious pitfalls.

Seeking to minimize the contamination by stars mistakenly classified
as galaxies and resulting in a false group detection, we restricted
our sample to CGs with galactic latitude $|b| > 40^{\circ}$, resulting
in a sample of 459 CGs distributed over 6260 square degrees of the
northern sky as shown in Figure 1. Around the north Galactic pole
(NGP) there are 352 CGs over 4637 square degrees, while in the
southern Galactic cap (SGC) region 107 CGs cover 1623 square degrees.
Both regions have similar surface densities of CGs, 0.0759 and 0.0659
groups per square degree, respectively, corroborating the homogeneity
of our sample over this large area. In comparison with the sample
presented in Iovino et al. (2003), there are 27 groups which are not
part of this newer sample due to rejection by the more stringent
galactic latitude cutoff (10 groups); improvement of bad areas defined
by bright stars and plate defects (see Gal et al. 2004, Lopes et
al. 2004) (8 groups); and discarding of a few additional plates which
did not meet our final quality standards (9 groups). The catalog
consists of 409 groups with 4 members, 47 groups with 5 members, and 3
groups with 6 members.

Table 1 lists the characteristic parameters for sixty CGs
representative of our sample. The columns are: (1) group name,
composed of PCG (Palomar Compact Group) with the RA and Dec
coordinates; (2,3) right ascension and declination (J2000) of the
group center; (4) radius of the group in arc minutes; (5) total
magnitude of the group, $\rm m_{gr}$, in $r$-band; (6) mean surface
brightness of the group in $r$-band; (7) magnitude interval between
the brightest and the faintest galaxy in the group, $\Delta{\rm
mag}_{\rm comp}$; (8) magnitude interval between the brightest group
member and the brightest outlier galaxy within the isolation radius,
$\Delta{\rm mag}_{\rm iso}$; and (9) the number of galaxies in the
group.

In Table 2, we list the photometric parameters for each galaxy in
these sixty groups in the same order as in Table 1. The columns are:
(1) group name, as in Table 1, plus a letter indicating each galaxy in
the group, ordered from brightest to faintest; (2,3) right ascension
and declination (J2000) for each galaxy in the group; (4) total
magnitude in $r$; (5) total color ($g-r$); (6) position angle, in
degrees, measured counter-clockwise; (7) ellipticity; and (8)
redshift, when it is available in the literature. We distinguish
between redshifts available through the SDSS and 2dF databases and
those coming from different sources in the NASA Extragalactic Database
(NED; we refer the reader to NED Sample Name Information,
http://nedwww.ipac.caltech.edu/samples/NEDmdb.html) as the former
constitute homogeneous data, and provide access to the spectra as
well. Taking the redshifts of all the galaxies available in this table
(155 from SDSS and 62 from NED) as representative of the whole sample
we estimate a median redshift of $z_{\rm med}$=0.12. The galaxies with
available spectra from SDSS are marked to identify those which can be
studied in more detail using public data. In Figure 2, we show the
DPOSS (F) images of the sixty CGs listed in Tables 1 and 2. The circle
indicates the group radius and the horizontal line indicates a length
of 0.5$^{\prime}$. The complete catalogs for Tables 1 and 2 with all
corresponding finding charts will be published electronically.

Before using this sample for any investigation of the physical nature
of CGs we need to understand the selection effects present, which, if
not taken into account, can make any conclusion misleading. We examine
the global features of this sample using the parameters $\Delta{\rm
mag}_{\rm comp}$, $\Delta{\rm mag}_{\rm iso}$, and $\mu_{\rm gr}$,
which were used in the sample definition, together with R$_{\rm
gr}$. The redshift information available for these groups is limited
(only 11\% of the galaxies, from 145 groups, have spectroscopic
redshifts), preventing more detailed analysis of the contamination
rate, mass estimates, and studies of dynamical and morphological
properties of these systems. However, the available photometric
parameters furnish a first glimpse into the reliability of the sample,
as representative of a physical class.

The distribution of the parameters noted above are shown in the four
panels of Figure 3. In panel (a) we show how the total magnitude of
the group varies with group radius. There is large scatter but a
majority of the sample is concentrated along the solid line indicating
the limiting surface brightness ($\mu_{\rm limit}$=24.0 mag
arcsec$^{-2}$) imposed as a selection criterion. Also, $\sim$70\% of
the sample is located between the solid and dashed lines (which
represents 1-$\sigma$ of $\theta_{\rm gr}$), a region where the points
follow the global trend indicated by the solid line. This trend
between total magnitude and size is expected for a sample with a
typical physical size and following the flux-distance
relation. However, we should emphasize that the concentration of
groups towards the solid line is due to the fact that setting a
surface brightness limit increases the probability of finding groups
with surface brightnesses closer to the border since high surface
brightness groups are rarer. Panel (b) shows $\mu_{\rm
gr}\times\Delta{\rm mag}_{\rm iso}$. We see a strong concentration of
low surface brightness systems, with 56\% of the sample having
$\mu_{\rm gr} > 23.5$. The vertical histogram shows that fainter
values of $\mu_{\rm gr}$ are preferred, a consequence of the higher
contamination rate for fainter $\mu_{\rm gr}$.  Panel (c) shows
$\Delta{\rm mag}_{\rm comp}\times\Delta{\rm mag}_{\rm iso}$, with the
solid line representing the criterion imposed by the algorithm
(${\Delta}\rm mag_{\rm iso}\geq \Delta \rm mag_{\rm comp}+0.5^{\rm
m}$).  Two effects must be taken into account when examining this
figure. First, groups will preferentially have $\Delta{\rm mag}_{\rm
comp}$ = 2, close to the criterion limit. Second, groups will tend to
have $\Delta{\rm mag}_{\rm iso}$ = 2.5, again near the border of the
selection criterion.  Finally, in panel (d) the relation between
$\mu_{\rm gr}$ and $\Delta{\rm mag}_{\rm comp}$ is plotted. No
correlation is present and the absence of groups in the lower left
corner of the diagram may indicate a selection effect due to the
cutoff in the magnitude of the brightest group member, 16$^{\rm m}$.
Galaxies in the magnitude interval between 16$^{\rm m}$ and 16.5$^{\rm
m}$ may be affected by saturation in the plate density measurement,
resulting in underestimation of the total magnitude. We note that at
these brighter magnitudes ($16\le \rm m_{\rm r}\le 17$), comparison
with the SDSS database has shown that we miss $\sim$14\% of these
bright galaxies due to non-detection and misclassification.

\subsection{The Space Density of Our Sample}

We estimate the space density of the CGs in our sample following the
method of Lee et al. (2004, their equation 1), using $z_{\rm med}$ =
0.12, resulting in a value of 1.6$\times$10$^{-5} h^{3}$
Mpc$^{-3}$. If we take the NGP and SGC areas separately we measure
space densities of 1.7$\times$10$^{-5} h^{3}$ Mpc$^{-3}$ and
1.5$\times$10$^{-5} h^{3}$ Mpc$^{-3}$, respectively, consistent with
the global measure. Lee et al. (2004) reports a value of
9.4$\times$10$^{-6} h^{3}$ Mpc$^{-3}$ when they apply their equation
to data from the Iovino et al. (2003) sample, while their estimate for
their own sample is 9.0$\times$10$^{-6} h^{3}$ Mpc$^{-3}$ when
matching their surface brightness criterion to ours. These estimates
are all fully consistent and indicate a space density of
$\sim1.0\times10^{-5} h^{3}$ Mpc$^{-3}$ for CGs in a slightly higher
redshift regime.

Mendes de Oliveira \& Hickson (1991) found a space density of
3.9$\times$10$^{-5} h^{3}$ Mpc$^{-3}$ for the nearby sample of Hickson
groups ($z_{\rm med}=0.03$), which is considerably higher than that
measured for the samples mentioned above. Barton et al. (1996) also
find a similar space density for nearby CGs, 3.8$\times$10$^{-5}
h^{3}$ Mpc$^{-3}$. These two estimates are consistent with each other
and were determined using totally different algorithms. If we force
the Hickson groups to satisfy our more stringent surface brightness
constraint, we find that their space density is reduced to 2.5
$\times$10$^{-5} h^{3}$ Mpc$^{-3}$. Also, a smaller $\Delta{\rm
  mag}_{\rm comp}$ would further reduce the space density. These
results suggest that local CGs have a higher space density compared to
their counterparts at higher redshifts, especially considering that
$\sim10\%$ contamination by projection effects may inflate our
estimate of the higher redshift space density.  However, as pointed
out by Lee et al. (2004) there are several possible explanations for
this discrepancy.  These samples are all based on different criteria
which can make such comparisons somewhat meaningless. Thus, it is
quite plausible that the discrepancies between nearby and more distant
samples reflect only differences in the way various authors defined
their search criteria.

\section{Groups \& Field}

In this Section, we investigate some of the group properties compared
to the field population, which is defined using the same galaxy
catalogs from DPOSS. For each plate catalog we discarded the CGs, as
well as galaxies within two Abell radii of clusters found in Gal et
al. (2003) and the bad areas where very bright stars heavily
contaminate the galaxy catalog (see Gal et al. 2002 for details). We
define the field population in the same plates used for the CG
catalog, resulting in a magnitude distribution where each bin
represents the median counts from all the plates used. Restricting
ourselves to the magnitude interval 16$<m_r<$19 (where group
galaxies are defined) we find that a linear fit to number counts
yields a slope of 0.49$\pm$0.01, which is consistent with that
obtained by Weir, Djorgovski, \& Fayyad (1995b), 0.52$\pm$0.01. This
agreement reinforces the characterization of our field population.

To examine whether or not groups are real or merely projection
effects, we compare their properties to those of simulated groups
composed of field galaxies. These aggregates are created by first
searching for relatively isolated galaxies with magnitudes within 0.1
mag of the brightest group galaxies. For each such field galaxy, we
then find the nearest four galaxies within 2 magnitudes, which is one
of the criteria for the construction of our real group sample. This
results in a sample of 36,000 fictitious groups whose properties can
statistically be compared to the actual group candidates.

The manner in which mergers proceed at different clustering scales is
a longstanding problem in cosmology. In a CG, we can think of
inelastic encounters leading to mergers (in some circumstances), and
as the cross section of the final product increases the later
accretions will occur more rapidly. In this scenario, CGs would evolve
into a single galaxy in as short as $\sim$10\% of the Hubble time, and
we would be faced with the dilemma that they are still observed. In
general, merging influences the luminosity function, and thus the
difference in brightness between the first and second (or later)
ranked galaxies, which can be assessed from statistical measures of
$\Delta{\rm mag}_{1-2}$. Numerical simulations have offered
interesting insights into this question. Mamon (1987) simulated
possible CG merger scenarios and found that $\Delta{\rm mag}_{1-2}$
increases very rapidly with the number of mergers, as we naively
expect, becoming quite large even before full coalescence is reached
(see his Figure 12). He then applies the Tremaine \& Richstone (1977)
test to measure how important mergers have been in these
systems. Below we describe the results of applying this test to
our sample. However, it is important to note that the evolution of
$\Delta{\rm mag}_{1-2}$ depends on several CG properties that are
currently unknown, including the initial conditions from which the
group originated, whether or not the CG resides in a single dark halo
or only contains the individual galaxy halos (Mamon 1987), and how
dissipation proceeds as the group is formed at an earlier epoch.

Figure 4 shows the distribution of the differences in magnitude
between the brightest and the second brightest galaxies ($\Delta{\rm
mag}_{1-2}$) in the real groups (solid line) compared to the
fictitious groups (dotted line). There is clearly a greater
occurrence of lower $\Delta{\rm mag}_{1-2}$ for galaxies in real
CGs. For the fictitious systems we find no relation between frequency
and $\Delta{\rm mag}_{1-2}$, as expected, since a priori these
galaxies do not interact amongst themselves. An additional test was
designed in order to better understand the distribution of $\Delta{\rm
mag}_{1-2}$.  Another fictitious sample was created by randomly
selecting galaxies from the actual magnitude distribution of galaxies
in real CGs. In this case we computed $\Delta{\rm mag}_{1-2}$ by
taking the brightest member galaxy for each of the quartets and a 
second galaxy randomly selected from the remaining three group members, 
with $\Delta{\rm mag}_{\rm comp}<$2. This procedure was repeated 20 times 
for the entire sample to improve the statistics. The dashed
line in Figure 4 indicates that the distribution of $\Delta{\rm
mag}_{1-2}$ for these fictitious groups differs from the one for real
groups. The fact that $\Delta{\rm mag}_{1-2}$ is smaller for real
groups than for the fictitious groups suggests that mergers may be the
primary mechanism for accreting small mass galaxies onto the most
massive ones. Tremaine \& Richstone (1977) proposed two measures as
evidence for mergers at the bright end, defined as T$_{1} = {\rm
rms}(M_{1})/\langle\Delta\rangle$ and T$_{2} = {\rm
rms}(\Delta)/(0.677)^{1/2}\langle\Delta\rangle$, where $\Delta$ is the
difference in magnitude between the brightest (M$_{1}$) and the second
brightest galaxy in the group. If mergers are important, then T$_{1}$
and T$_{2}<$1. Mamon (1986) finds T$_{1}$=1.16 for 41 Hickson groups
with concordant redshifts and suggests that there is no evidence for
luminosity function evolution in the Hickson sample. Using the 92
concordant redshift Hickson groups, which include triplets, we find
T$_{2}$=0.99, corroborating Mamon's result. In our case, since we do
not have redshifts to measure T$_{1}$ properly, we have used T$_{2}$
instead, and find a value of 0.90$\pm0.12$, indicating possible
mergers for the bright galaxies in our groups. The error quoted for
T$_{2}$ comes from bootstrapping the sample 1000 times for 90\% of the
groups, as we expect 10\% contamination that might affect the $\Delta$
measurements. We emphasize that this result is merely suggestive, but
in combination with other evidence, such as the differing
distributions of $\Delta{\rm mag}_{1-2}$ for groups and the field,
implies that at least part of this sample should be representative of
real structures and not result from projection effects.

Another important piece of information comes from the distribution of
$g-r$ colors of the group galaxies. In Figure 5 we present the
histograms of colors for galaxies in CGs (solid line) and for field
galaxies, where we note a clear excess (5-$\sigma$ considering
Poissonian errors) of the former in the interval $0.2 < (g-r)<
0.5$. Two important points should be noted here: first, the $g-r$ color
does not distinguish between early and late type galaxies in the
redshift regime of this sample, so we cannot conclude anything
about morphological dominance in groups; second, from the figure we
can see that the FWHM of the color distribution for group galaxies is
approximately a factor of two smaller than the one for field
galaxies. Therefore, there seems to be an excess in the central bins of
the distribution indicating that there is more color concordance among
CG galaxies than among field galaxies, strengthening our view that this
sample of groups represents physical entities and not projection
effects.

\section{Groups and Large Scale Structure}

Although little work has been done on modeling and simulating the
formation and evolution of compact groups, in the paradigm of
hierarchical structure formation we might expect that they form in
enhanced density regions and are therefore related to other more
significant structures like filaments and clusters (e.g. West 1989,
Hernquist, Katz, \& Weinberg 1995). In what follows when we refer to the
''relation'' between CGs and the LSS we do not attempt
any precise definition of it. We basically use angular proximity as
a first indicator. In order to quantify the degree of association on
firmer grounds we would need a spectroscopic survey complete down to a
limiting magnitude that allows a consistent probe of the luminosity
function of all observed structures.

Several observational papers addressed the density distribution in the
environments of CGs, which is a first approximation of the relation
between CGs and other structures in the universe (e.g. Sulentic 1987,
Palumbo et al. 1995). They conclude that most of the CGs cannot result
from chance alignments. However, the samples available at that time
and the subjective nature of their measurements hampered any more
serious investigation into how CGs are distributed relative to the
LSS. Rood \& Struble (1994) and Ramella et al. (1994), following
independent ways of establishing such a relation, found that CGs are
often embedded in sparser structures. These results were confined to
the Hickson sample. More recently, Andernach \& Coziol (2004) used the
sample of 84 CGs from Iovino et al. (2003) and the cluster sample from
Gal et al. (2003) to address this issue and found that 52\% of the
PCGs in Iovino's sample are related to clusters. With our sample we
can once again examine the CG-LSS association. First, we compare our
sample with other catalogs available in the literature through
NED. This matching also allows us to show how our automatic search is
able to reproduce previous visual searches. Second, we examine the
relation between our CG catalog and that of galaxy clusters from Gal
et al. (2003).

We performed a preliminary cross-match between our sample and those
available in the literature to examine the relation between the groups
defined by our objective search and other structures present in the
universe, most of which were detected in a subjective manner. Again it
is important to bear in mind that association here means angular
proximity.  Data were collected from NED.  We searched for structures
including triplets, groups, and clusters within 16.5$'$ of our CGs,
which corresponds to 1.5 $h^{-1}$Mpc at our estimated median redshift
$z_{\rm med} = 0.12$.  We found that 303 CG candidates (66\% of the
total sample) are close to at least one other structure, with most of
the associations to clusters (69\%). Additionally, 28\% are found near
groups and 3\% near triplets. This indicates that the CGs in our
sample are to some degree associated with the large scale structure,
which is not a surprising result but requires more quantitative
assessment. Besides of demonstrating a connection between CGs and the
LSS, this comparison has proven valuable in recovering groups
previously found by other searches (most of them visually).  If we
take all the clusters associated with groups and assume that both are
at the same redshift, we find a median redshift of $z_{\rm med}$ =
0.14, consistent with the estimate of Iovino et al. (2003) and our
estimate for those groups with at least one measured redshift (see
Table 2). This makes our catalog the largest and deepest objectively
composed sample to date, and together with the work of Lee et
al. (2004) based on the SDSS, forms the basis for further studies of
this clustering scale.

Considering that NED is a compilation of a wide variety of sources
with no homogenization, we require a more homogeneous database to
avoid inconsistencies in the way we establish a relation between the
CGs and the LSS.  The recently published list of galaxy clusters from
DPOSS by Gal et al. (2003) provides the most appropriate comparison
between CGs in our sample and clusters of galaxies. First, it is drawn
from the same galaxy catalog; second, it covers a significant fraction
of the northern sky; and finally, it provides important information
needed for this comparison, including photometric redshifts and
richnesses. We note that Gal et al. (2003) covers only the
well-calibrated NGP area to which we restrict our comparison. For each
of the 352 CGs in the NGP area we determine the distance to the
nearest cluster, expressed as the number of Abell radii at the cluster
redshift, k$_{\rm Abell}$, shown by the vertical solid line in all
panels of Figure 6. Panel (a) shows the distribution of k$_{\rm
Abell}$ as the solid line. We also estimate the distribution of
distances to an equal number of points chosen randomly in the same sky
region. We repeat this experiment 1000 times, and plot the median
distribution of these distances as the dashed line in Figure 6. As a
further test, we measure the distance between bright galaxies
(16.0$<m_r<$17.0) and clusters, selecting 352 galaxies from the NGP
area and repeating this search 1000 times. This distance distribution
is shown as the dotted lines in Figure 6. All three distributions are
normalized by the number of points. As we expect, the distribution of
k$_{\rm Abell}$ for random and bright galaxies are similar, confirming
that the strong spatial coincidence between CGs and clusters is not
due to our group selection algorithm.

When comparing with all clusters, there is an excess of groups within
one Abell radius over the random distribution (32\%), in contrast to
the percentage we found when comparing to the NED database
(66\%). Moreover, by dividing the cluster sample by the median
richness, we can examine how CGs are related to low and high mass
environments, assuming there is a relation between richness and mass
(see Bahcall et al. 2003).  In panel (b) of Figure 6 we plot the
distribution of angular separations between CGs and rich clusters,
where only 20\% (over random) of the CGs are within one Abell radius,
while in panel (c), for poor clusters this percentage drops to
13\%. Therefore, we find a marginal excess of CGs related to rich
clusters relative to poor clusters. The large excess of small
separations in the bright galaxy distribution is due to luminosity
segregation in clusters. The important result is that both
distributions agree extremely well for ${\rm k_{Abell}}\ge 3.0$. These
results uniquely display how CGs are associated with large scale
structure. The fact that the distributions for the groups shown in
Figure 6 are similar to the expectation for random samples and bright
galaxies at ${\rm k_{Abell}}\ge 2.5$ shows that a significant fraction
of CGs resides in the low density regions. This is an important clue
for scenarios of galaxy formation and evolution, since most of the CGs
in our sample do not seem to be associated to either rich or poor
clusters.

Using our galaxy database we applied a simple statistical test to
study the environment of our CG candidates.  As a first test we
counted for each group of our sample the density of galaxies
($\rho_{in}$) within $\rm R_{\rm gr}$ in the magnitude range $\rm
m_{\rm brightest} < m < m_{\rm brightest} + \Delta \rm mag_{\rm comp}
+1.5 $ and compared it to the density of galaxies ($\rho_{\rm out}$)
in the same magnitude range but in the ring defined by $3\rm R_{\rm
gr} < \rm R < 15\rm R_{\rm gr}$.  Panel (a) of Figure 7 shows the
distribution of the ratio $\rho_{\rm in}/\rho_{\rm out}$, or, in other
words, the density contrast of CGs in our sample with respect to
their immediate neighborhood (having excluded the isolation ring). The
vast majority of groups have a density contrast $\rho_{\rm
in}/\rho_{\rm out}$ greater than 10 and $\sim 50\%$ greater than
35. This result suggests that CGs in our samples are large
overdensities with respect to the local environment, even when
disregarding the isolation ring (that, by definition, is empty of
galaxies of magnitude comparable to those of the group members).

But how does the local environment of our CG candidates compare to
that of a generic galaxy in the database? To answer this question we
compared $\rho_{out}$, as defined above, with the same quantity for a
generic field galaxy in the database.  We considered, for each
CG, all galaxies in the database with magnitude within the range
$\rm m_{\rm brightest}-0.1^{\rm m} < \rm m < \rm m_{\rm
brightest}+0.1^{\rm m}$ and measured the surrounding density using the
same prescriptions adopted to measure $\rho_{\rm out}$ for the
group. We then calculated for each group the mean value $\rho_{\rm
field}$ and its scatter $\sigma_{\rm field}$. Panel (b) of Figure 7 shows the
histogram of the quantity $(\rho_{\rm out}-\rho_{\rm field})/\sigma_{\rm field}$.
The distribution shows that the vast majority of groups in our sample
inhabit regions of density undistinguished from those of the
background field distribution.  The small tail of groups at values
$(\rho_{out}-\rho_{field})/\sigma_{\rm field} > 3$ are possibly the few
subcondensations within larger structures surviving the selection
criteria.

Another important correlation is shown in Figure 8, where we
investigate how three important group properties, total magnitude
($\rm m_{gr}$), surface brightness ($\mu_{\rm gr}$), and radius, vary
with the distance to the nearest cluster.  Data were binned keeping
the same number of points in each bin (20), and the error bars are the
quartiles of the distribution. We compute the correlation coefficients
s (Spearman) and k (Kendall) to show how the parameters are correlated
to the distance to the nearest cluster. In principle, if CGs form a
truly independent family we should see no correlations. Both correlation 
coefficients shown in panel (a) indicate a very weak dependence of group
surface brightness on k$_{\rm Abell}$. Panel (a) also
shows that CGs further away from clusters have low surface
brightnesses, while those close to clusters cover the entire range in
$\mu_{\rm gr}$ with a predominance of low surface brightness CGs. This
may indicate that at least a fraction of CGs close enough to clusters
might be part of the larger structure responding to the global
potential instead of a dynamically independent system. In other words,
for k$_{\rm Abell}<$1.0 the sample could be heavily affected by
projection effects. Both correlation coefficients indicate a very weak
dependence of group surface brightness on k$_{\rm Abell}$.  Panel (b)
shows no variation of the total magnitude of CGs with k$_{\rm Abell}$,
demonstrating how difficult it is for an objective algorithm to
distinguish between a truly isolated CG and one in the center of a
larger structure.  Panel (c) shows that the groups' radii do not vary
significantly either. Although in this comparison between both samples (CGs
and clusters) there is a strong assumption that CGs are at the same
redshift as the angularly nearest cluster, this result suggests that
truly compact and isolated systems, as idealized by Hickson, should be
found in very low density regimes, although we cannot rule out the
possibility that at least a fraction of them reside in sparse
groups. Spectroscopic follow-up will be essential to conclusively
demonstrate these results.

\section{Summary}

Searching for small groups over an area of 6260 square degrees around
the north and south Galactic caps, using the DPOSS galaxy catalog, we
have found 459 CGs expected to be at $z_{\rm med}\sim$0.12 and
extending to $z = 0.2$. As shown in Iovino et al. (2003) the
contamination rate is $\sim$10\%, taking into account only projection
effects. Although we apply a fairly high cutoff in galactic latitude
(40$^{\circ}$), it is possible that some contamination by stars is
still present. This will have to be verified by spectroscopic
follow-up, which will allow us not only to remove chance alignment
systems but also to address important questions regarding how galaxies
evolve in different density regimes.

Although the redshift information available for our sample is not
sufficient for a proper dynamical analysis, we are able to address two
important questions regarding the nature of CGs. First, we find that
the space density of the CGs in our sample (1.6$\times$10$^{-5} h^{3}$
Mpc$^{-3}$) is very similar to that obtained by Lee et al. (2004) for
the SDSS CG sample (9.4$\times$10$^{-6} h^{3}$ Mpc$^{-3}$) when
applying a similar surface brightness limit. Second, we find that CGs
in our sample are associated with clusters at a level of 32\%, with a
marginal tendency to be more related to rich clusters. This level of
association is much lower than that found by previous studies for
nearby CGs, such as 75\% in Rood \& Struble (1994). However, this
discrepancy can be due to several reasons, including different ways of
establishing the association between CGs and clusters.  Rood \&
Struble (1994) study the association of CGs with not only clusters but
also sparse groups, and since these systems dominate the large scale
structure in the Universe (Nolthenius \& White 1987) it is
unsurprising to find a higher rate of association if they are
included. Finally, there is no a priori reason to obtain a similar
degree of association as we move out in redshift.

Galaxy evolution is already detected at intermediate redshifts 
$z\sim$0.1-0.2 (the Butcher-Oemler effect, e.g. Butcher \& Oemler
1978, Margoniner \& de Carvalho 2000, Carlberg et al. 2001). With this
objectively selected sample we will be able to establish a firm
comparison between predictions and observations, which is one of
the most problematic issues in the study of compact groups.

\acknowledgments We thank the Norris Foundation for their generous
support of the DPOSS project and the creation of the Palomar-Norris
Sky Catalog (PNSC).  We also thank the Palomar TAC and Directors for
generous time allocations for the DPOSS calibration effort. RRdC would
like to thank Gary Mamon and Steve Zepf for several useful discussions
throughout this project. SGD acknowledges a partial support from the
Ajax Foundation. This work was made possible in part through the NPACI
sponsored Digital Sky project and a generous equipment grant from SUN
Microsystems. Access to the POSS-II image data stored on the HPSS,
located at the California Institute of Technology, was provided by the
Center for Advanced Computing Research.  This research has made use of
the NASA/IPAC Extragalactic Database (NED) which is operated by the
Jet Propulsion Laboratory, California Institute of Technology, under
contract with the National Aeronautics and Space Administration.

\clearpage

\begin{figure}
\plotone{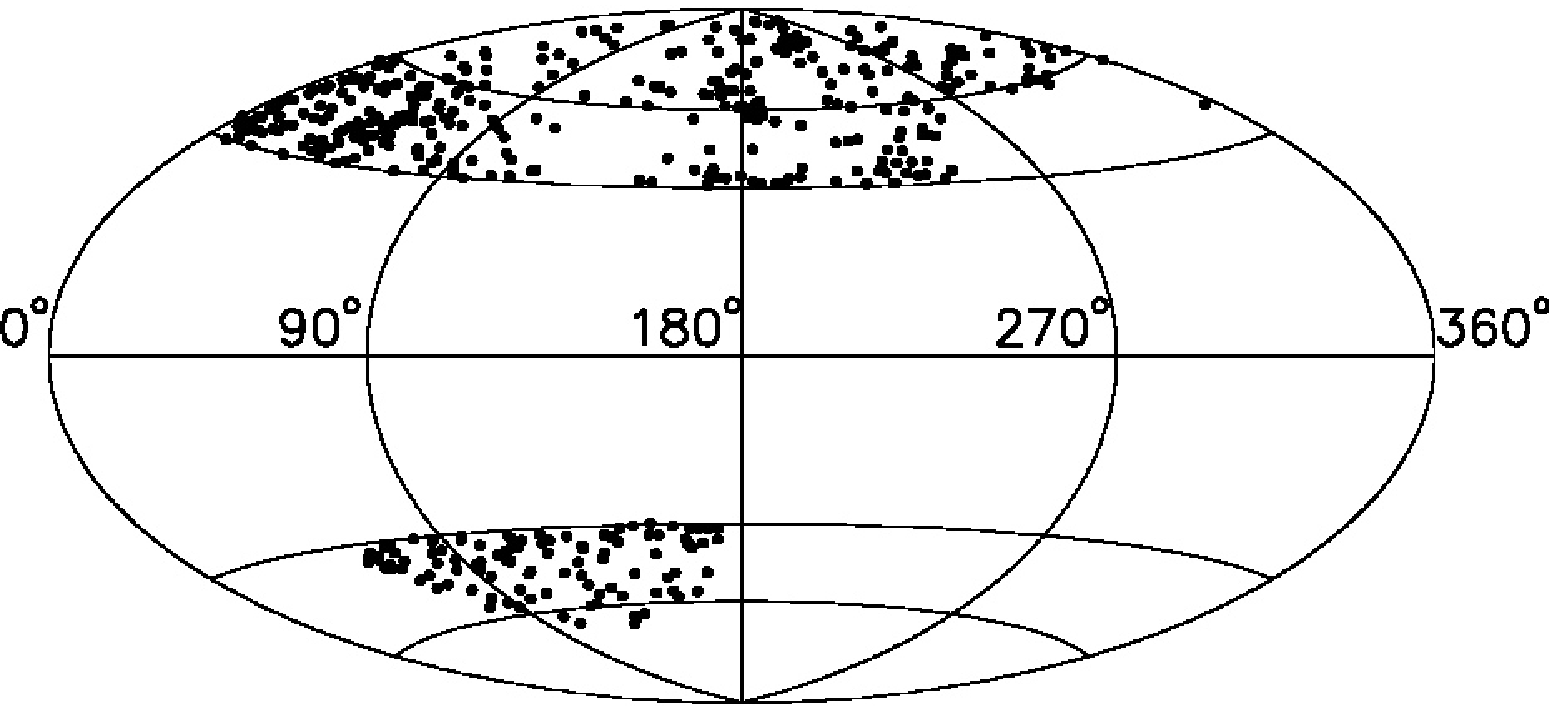}
\caption{Distribution of the 459 compact groups of our sample in
galactic coordinates.  The solid line indicating the limit of
40$^{\circ}$ in galactic latitude is clearly seen.}
\label{Figure 1}
\end{figure}

\clearpage

\begin{figure}
\vskip -1.5 truecm
\plotone{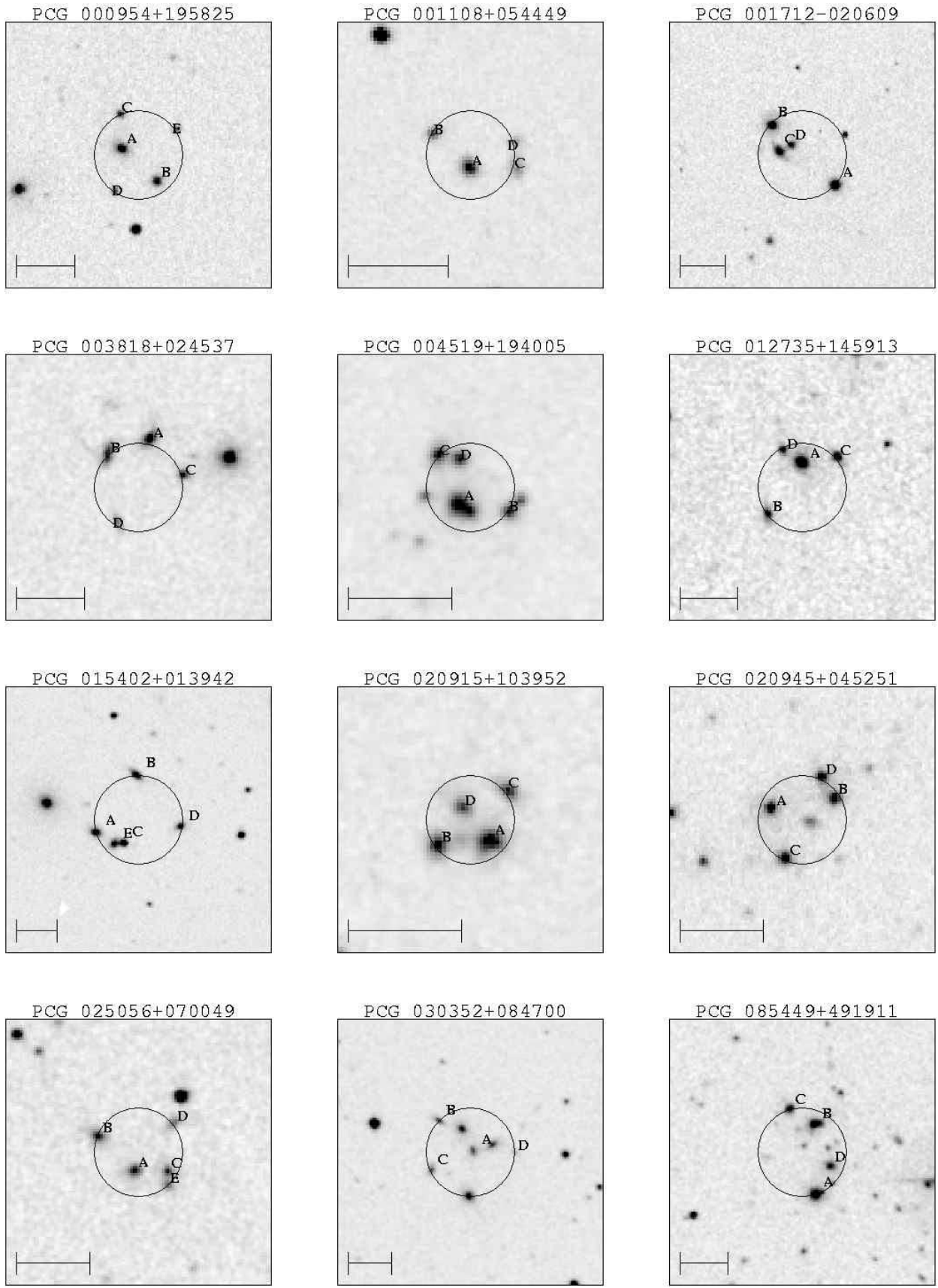}
\caption{DPOSS (F) images of 60 representative compact group
candidates in our sample. The horizontal bar indicates
0.5$^{\prime}$.}
\label{Figure 2}
\end{figure}
\clearpage

\setcounter{figure}{1}
\begin{figure}
\plotone{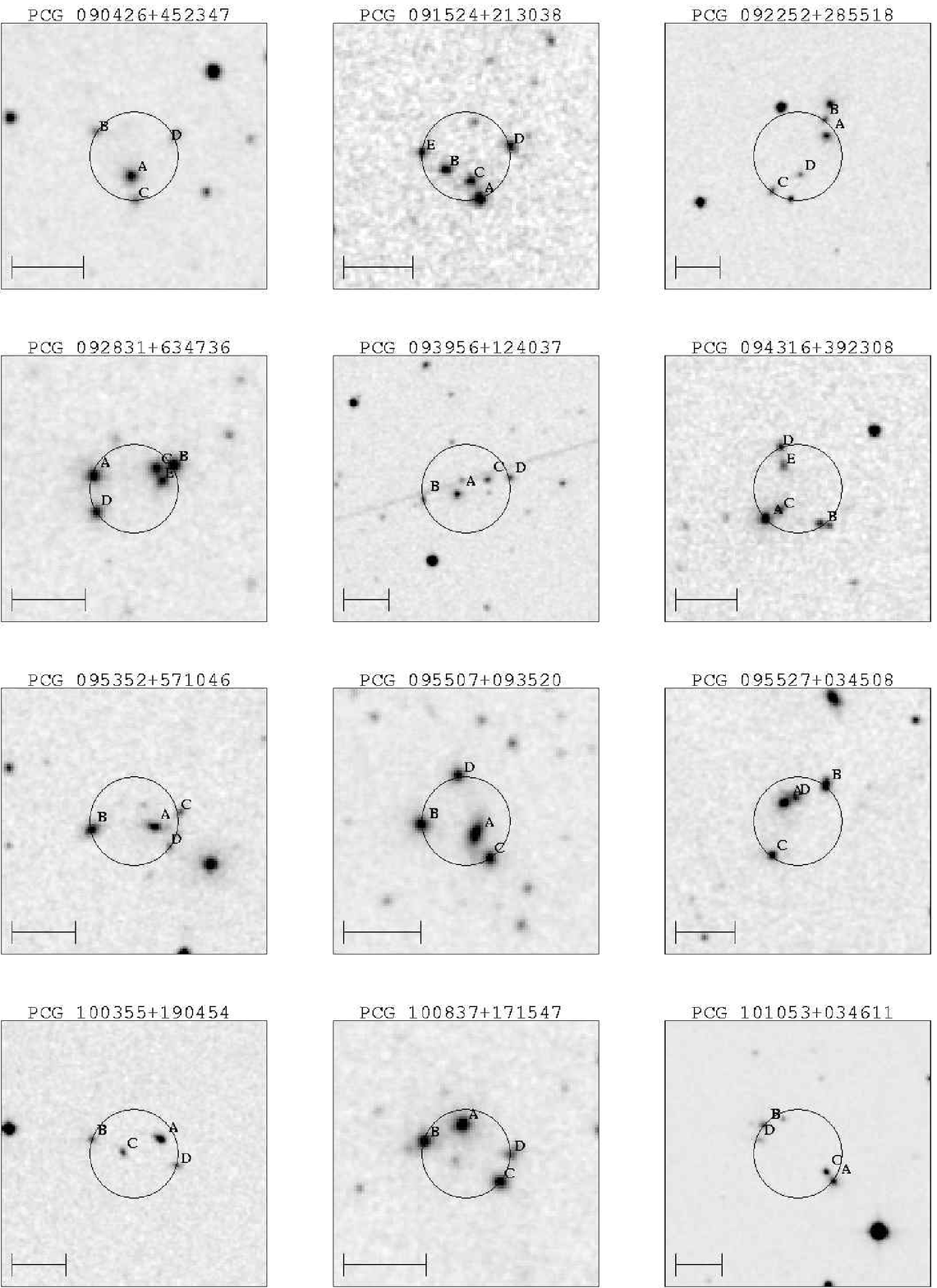}
\caption{Continued}
\label{Figure 2}
\end{figure}
\clearpage

\setcounter{figure}{1}
\begin{figure}
\plotone{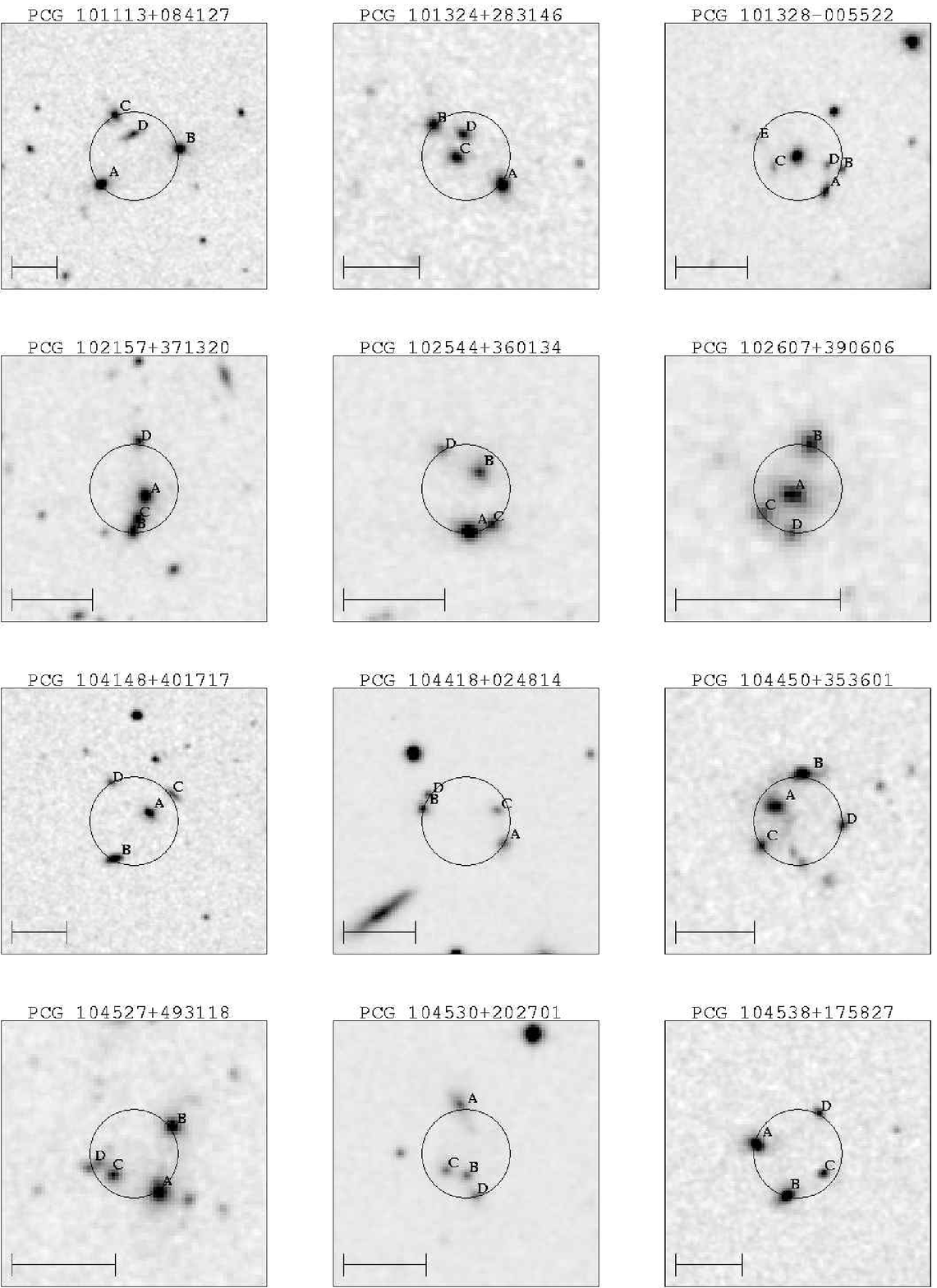}
\caption{Continued}
\label{Figure 2}
\end{figure}
\clearpage

\setcounter{figure}{1}
\begin{figure}
\plotone{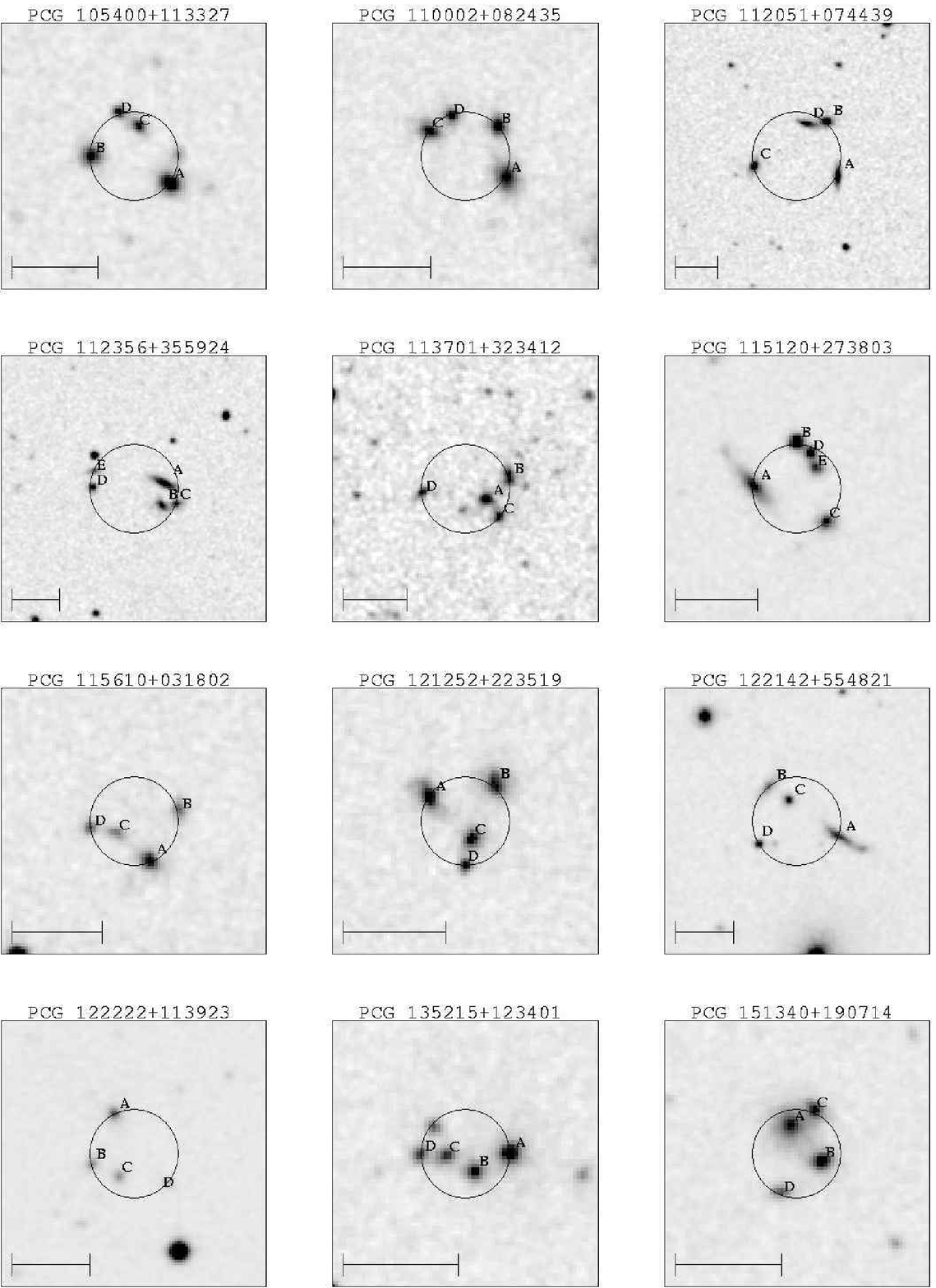}
\caption{Continued}
\label{Figure 2}
\end{figure}
\clearpage

\setcounter{figure}{1}
\begin{figure}
\plotone{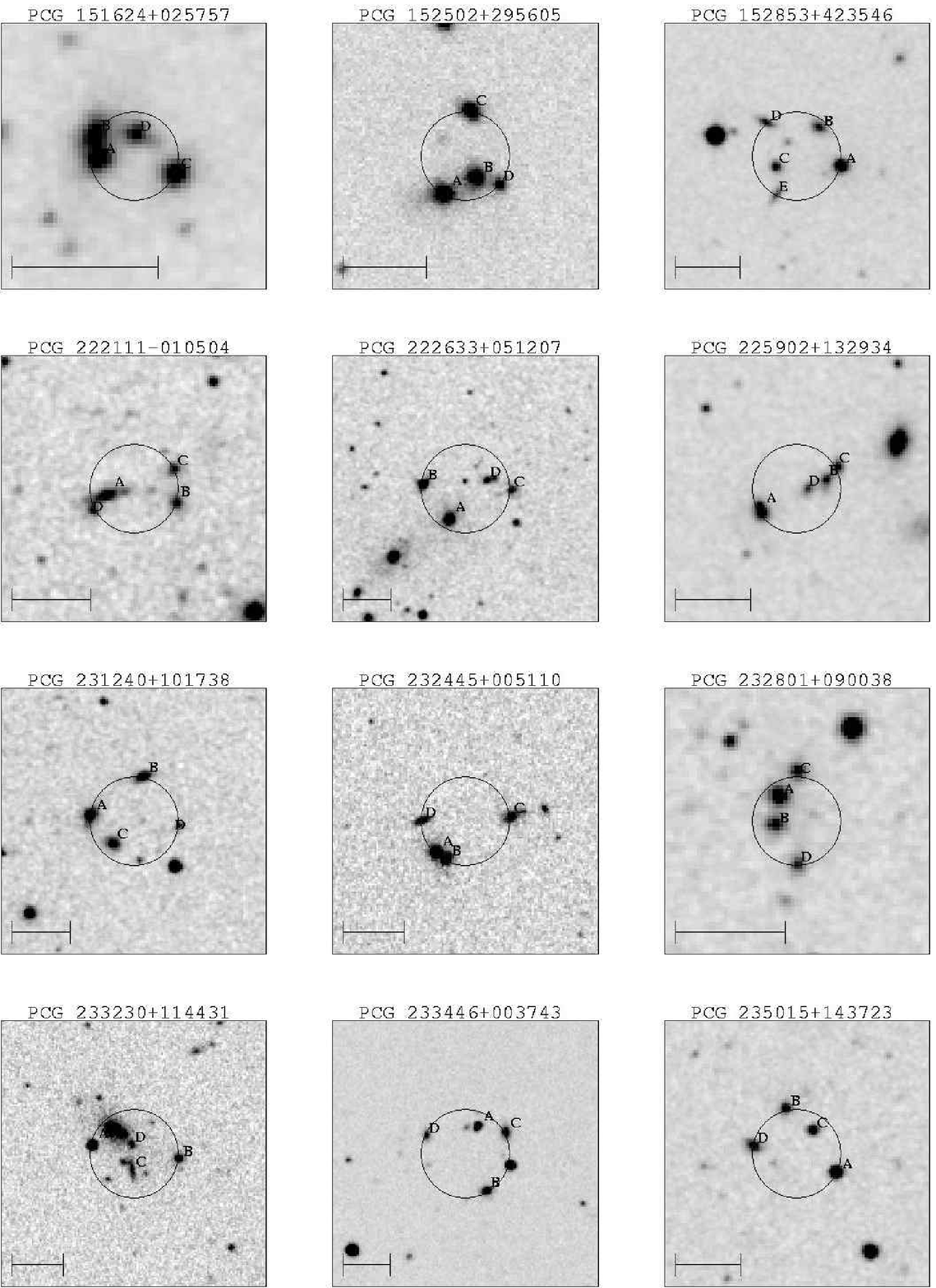}
\caption{Continued}
\label{Figure 2}
\end{figure}
\clearpage

\begin{figure}
\plotone{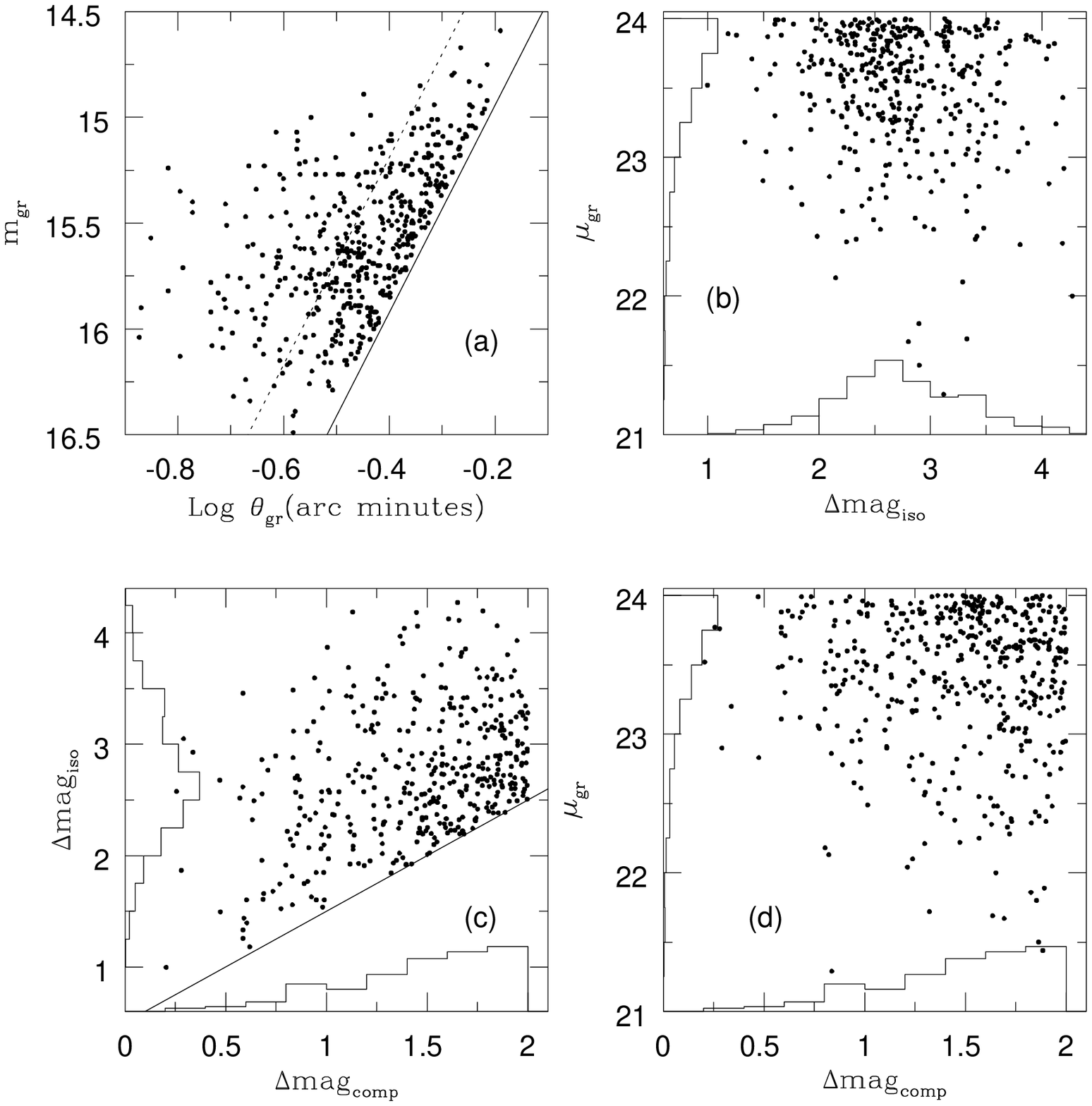}
\caption{Global properties of the 459 groups in our
sample. Correlations between $\rm m_{gr}$, $\mu_{\rm gr}$, $\Delta \rm
mag_{\rm comp}$, $\Delta \rm mag_{\rm iso}$, and $\theta_{\rm gr}$
display the limitations of the present sample so that comparison with
models are meaningful. In panel (a), the solid line indicates the
limiting surface brightness ($\mu_{\rm limit}$=24.0 mag arcsec$^{-2}$)
imposed as a selection criterion, and together with the dashed line
encompasses a region where 70\% of the points lie (width of 1-$\sigma$
of $\theta_{\rm gr}$). In panel (c), the solid line represents the
criterion ${\Delta}\rm mag_{\rm iso}\geq \Delta \rm mag_{\rm
comp}+0.5^{\rm m}$ (see text for details).}
\label{Figure 3}
\end{figure}
\clearpage

\begin{figure}
\plotone{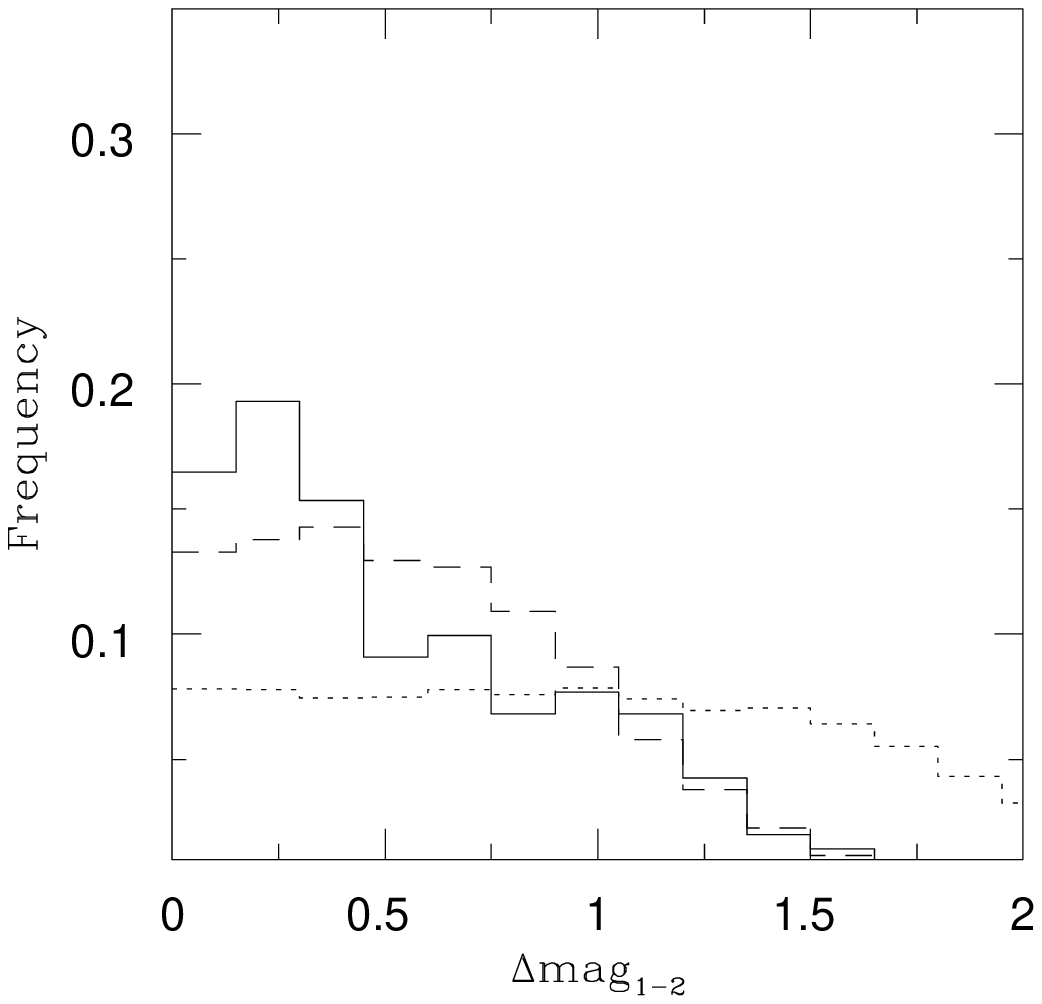}
\caption{Distributions of $\Delta{\rm mag}_{1-2}$ for CGs in our
sample (solid line) and the fictitious groups defined from the field
population (dotted line) and fictitious groups following the 
magnitude distribution of the real groups (dashed line). All
distributions are normalized to unity.}
\label{Figure 4}
\end{figure}
\clearpage

\begin{figure}
\plotone{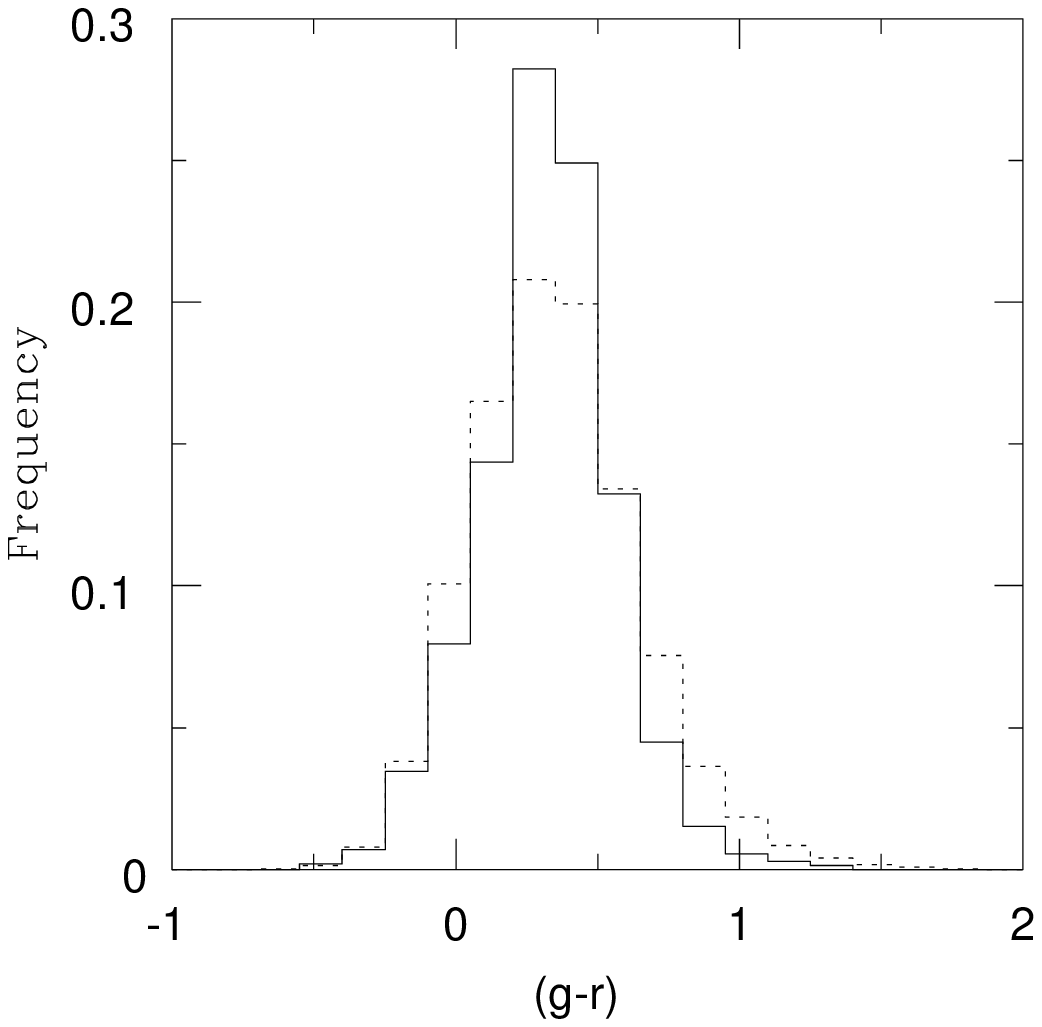}
\caption{Distributions of ($g-r$) color for CGs in our sample (solid line) and the fictitious
groups representing the field population (dotted line).}
\label{Figure 5}
\end{figure}
\clearpage

\begin{figure}
\plotone{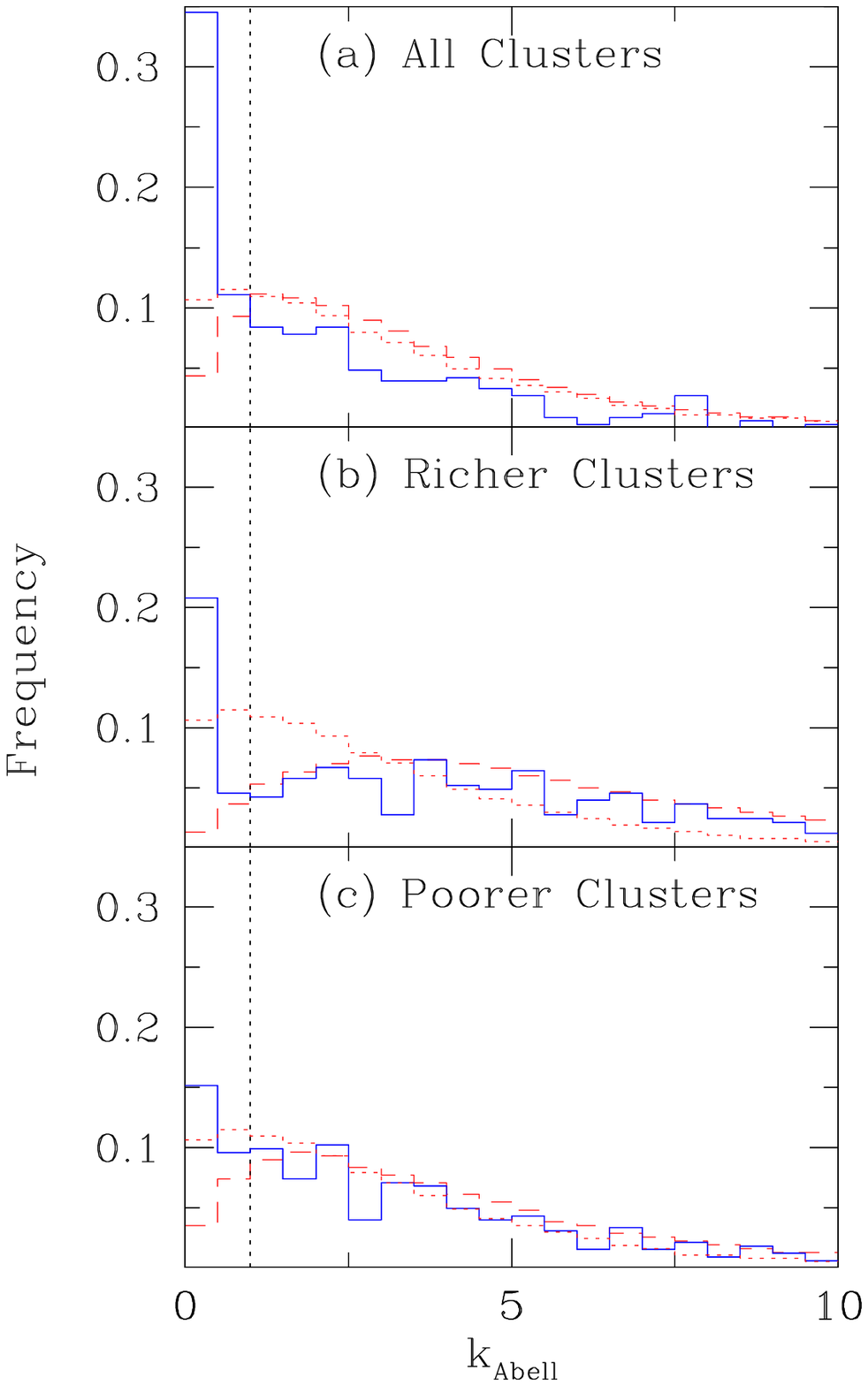}
\caption{Distributions of the distance to the nearest cluster. As
indicated, panels (a), (b), and (c) show the distributions using all
clusters in Gal et al. (2003), only rich clusters, and only poor
clusters, respectively.  Real groups are plotted as the solid line,
while the random distributions and the bright galaxies distributions
are represented by dashed and dotted lines, respectively.}
\label{Figure 6}
\end{figure}
\clearpage

\begin{figure}
\plotone{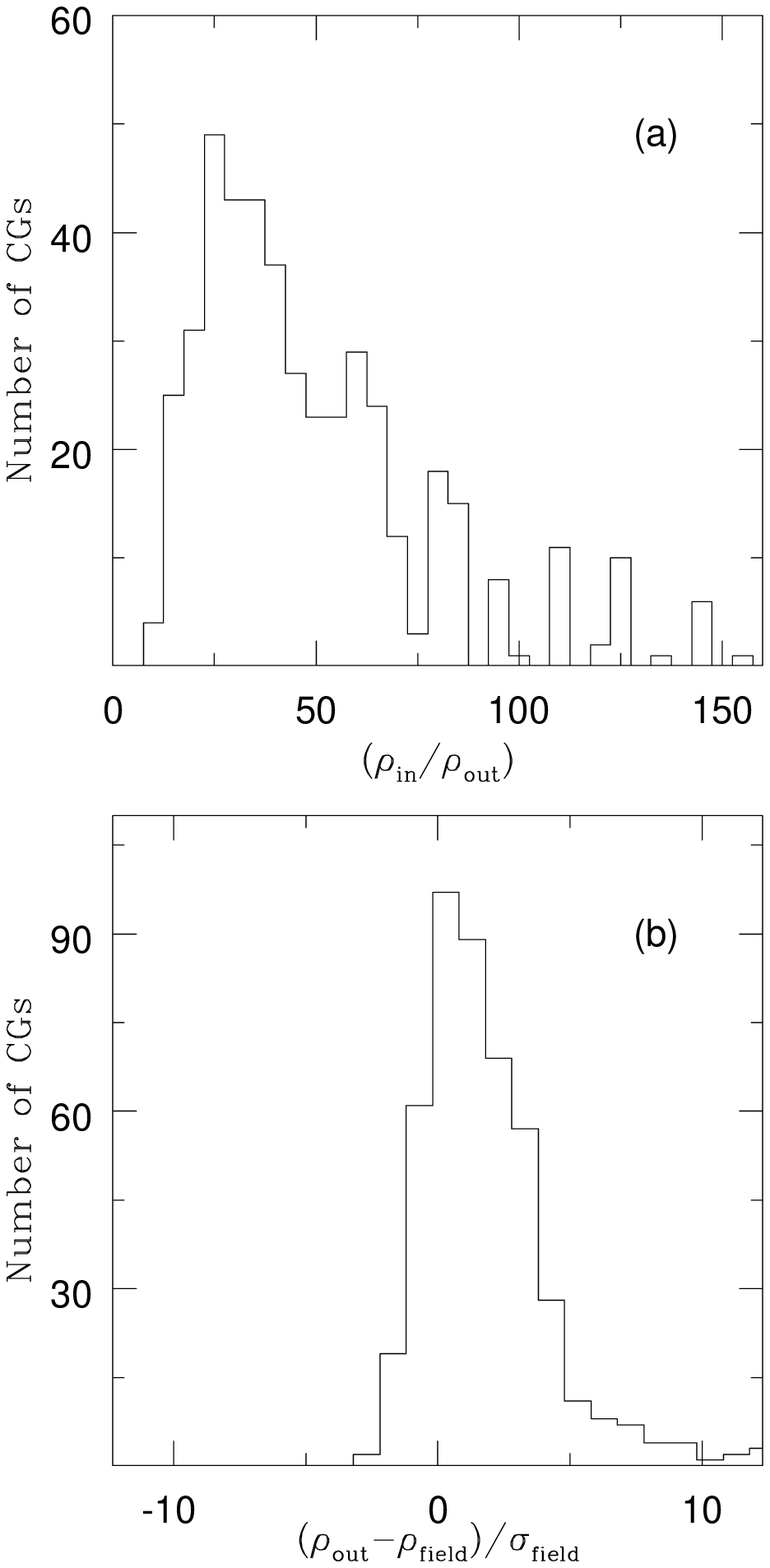}
\caption{Panel (a) exhibits the density contrast of the CGs in our
sample {\sl wrt} to the immediate neighborhood, while in panel (b) we show
how the local environment of our CG candidates compares to that of a
generic field galaxy, through the quantity $(\rho_{\rm out}-\rho_{\rm
field})/\sigma_{\rm field}$ (see text for details).}
\label{Figure 7}
\end{figure}
\clearpage

\begin{figure}
\plotone{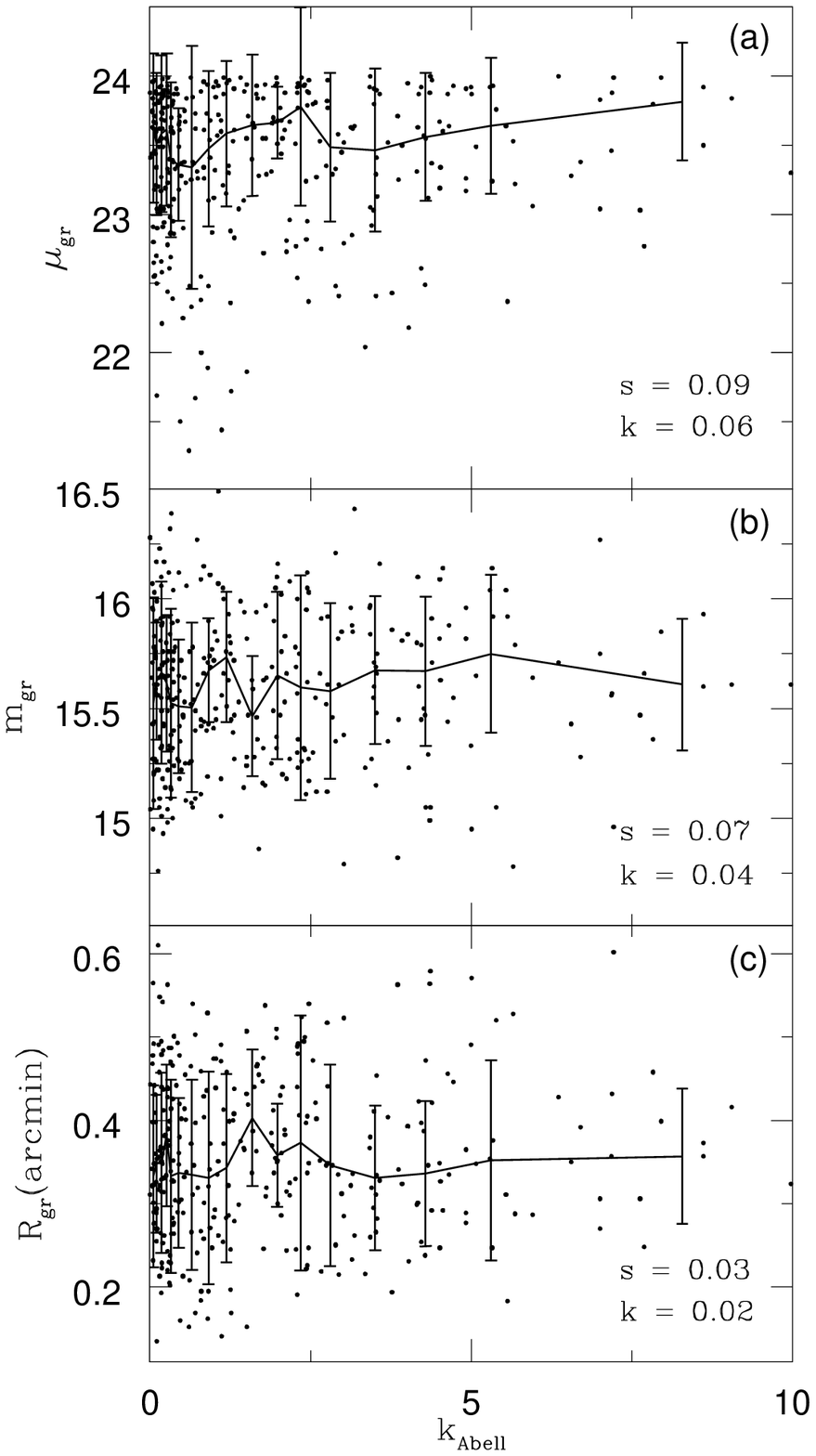}
\caption{The correlation between group surface brightness (Panel a),
total magnitude (Panel b), and radius (Panel c), and the distance to the
nearest cluster (k$_{\rm Abell}$). Spearman rank (s) and Kendal (k) correlation
coefficients are shown in each panel.}
\label{Figure 8}
\end{figure}
\clearpage

\begin{deluxetable}{ccccccccccc}
\tabletypesize{\scriptsize}
\tablecaption{Characteristic Parameters of the Groups in our Sample}
\tablewidth{0pt}
\tablehead{
\colhead{Name} &\colhead{RA} &\colhead{DEC}& \colhead{R$_{\rm gr}$} & \colhead{m$_{\rm gr}$}
& \colhead{$\mu_{\rm gr}$}& \colhead{$\Delta \rm mag_{\rm comp}$}& \colhead{$\Delta \rm mag_{\rm isol}$}&Number&
\\
\colhead{}&  \colhead{(J2000)}&  \colhead{(J2000)}     &  \colhead{arcmin}      & \colhead{mag}
&\colhead{$(\rm mag arcsec^{-2})$}&\colhead{mag}&\colhead{mag}&\colhead{}&
}
\startdata
PCG0009+1958 & 00 09 54.48 & +19 58 25.03 & 0.380 & 15.95 & 23.98 & 1.834 & 3.149 & 5 & \\[1pt]
PCG0011+0544 & 00 11  8.97 & +05 44 49.13 & 0.222 & 16.08 & 22.95 & 1.819 &   -   & 4 & \\[1pt]
PCG0017-0206 & 00 17 12.53 & -02 06 09.22 & 0.493 & 15.13 & 23.73 & 1.160 & 1.931 & 4 & \\[1pt]
PCG0038+0245 & 00 38 18.24 & +02 45 37.62 & 0.325 & 16.00 & 23.69 & 1.673 & 2.791 & 4 & \\[1pt]
PCG0045+1940 & 00 45 19.32 & +19 40 05.99 & 0.215 & 15.80 & 22.60 & 1.398 &   -   & 4 & \\[1pt]
PCG0127+1459 & 01 27 35.14 & +14 59 13.88 & 0.388 & 15.84 & 23.92 & 1.558 & 2.524 & 4 & \\[1pt]
PCG0209+1039 & 02 09 15.12 & +10 39 52.45 & 0.196 & 15.51 & 22.10 & 1.238 & 3.289 & 4 & \\[1pt]
PCG0209+0452 & 02 09 45.12 & +04 52 51.24 & 0.267 & 15.56 & 22.83 & 0.474 & 1.496 & 4 & \\[1pt]
PCG0250+0700 & 02 50 56.96 & +07 00 49.43 & 0.303 & 15.62 & 23.16 & 1.237 & 2.612 & 5 & \\[1pt]
PCG0303+0847 & 03 03 52.37 & +08 47 00.60 & 0.515 & 15.13 & 23.82 & 1.874 & 2.616 & 4 & \\[1pt]
PCG0854+4919 & 08 54 49.18 & +49 19 11.93 & 0.468 & 15.27 & 23.75 & 1.119 & 2.282 & 4 & \\[1pt]
PCG0904+4523 & 09 04 26.79 & +45 23 47.26 & 0.311 & 16.28 & 23.88 & 1.618 & 2.228 & 4 & \\[1pt]
PCG0915+2130 & 09 15 24.57 & +21 30 38.81 & 0.324 & 15.81 & 23.50 & 1.011 & 2.588 & 5 & \\[1pt]
PCG0922+2855 & 09 22 52.71 & +28 55 18.37 & 0.500 & 15.31 & 23.94 & 1.226 & 3.170 & 4 & \\[1pt]
PCG0928+6347 & 09 28 31.26 & +63 47 36.10 & 0.301 & 15.27 & 22.80 & 0.977 & 2.574 & 5 & \\[1pt]
PCG0939+1240 & 09 39 56.17 & +12 40 37.70 & 0.491 & 15.01 & 23.60 & 0.831 & 1.558 & 4 & \\[1pt]
PCG0943+3923 & 09 43 16.67 & +39 23  8.41 & 0.366 & 16.05 & 24.00 & 1.402 &   -   & 5 & \\[1pt]
PCG0953+5710 & 09 53 52.68 & +57 10 47.31 & 0.350 & 15.43 & 23.28 & 1.842 & 3.216 & 4 & \\[1pt]
PCG0955+0935 & 09 55  7.57 & +09 35 20.58 & 0.290 & 15.20 & 22.65 & 0.946 & 2.382 & 4 & \\[1pt]
PCG0955+0345 & 09 55 27.22 & +03 45  8.35 & 0.376 & 15.22 & 23.23 & 1.181 & 3.347 & 4 & \\[1pt]
PCG1003+1904 & 10 03 55.26 & +19 04 54.66 & 0.411 & 15.60 & 23.80 & 1.453 & 4.062 & 4 & \\[1pt]
PCG1008+1715 & 10 08 37.84 & +17 15 47.59 & 0.269 & 15.22 & 22.50 & 1.366 & 2.946 & 4 & \\[1pt]
PCG1010+0346 & 10 10 53.65 & +03 46 12.90 & 0.480 & 15.23 & 23.77 & 1.235 & 2.453 & 4 & \\[1pt]
PCG1011+0841 & 10 11 13.40 & +08 41 27.24 & 0.492 & 15.08 & 23.67 & 0.950 & 1.602 & 4 & \\[1pt]
PCG1013+2831 & 10 13 24.71 & +28 31 47.06 & 0.295 & 15.55 & 23.03 & 1.100 & 3.434 & 4 & \\[1pt]
PCG1013-0055 & 10 13 28.73 & -00 55 22.01 & 0.309 & 15.97 & 23.55 & 1.601 & 2.203 & 5 & \\[1pt]
PCG1021+3713 & 10 21 57.32 & +37 13 20.49 & 0.277 & 15.76 & 23.11 & 1.123 & 3.428 & 4 & \\[1pt]
PCG1025+3601 & 10 25 44.54 & +36 01 34.78 & 0.220 & 15.59 & 22.44 & 1.847 &   -   & 4 & \\[1pt]
PCG1026+3906 & 10 26  7.76 & +39 06  6.37 & 0.135 & 15.90 & 21.69 & 1.634 & 3.327 & 4 & \\[1pt]
PCG1041+4017 & 10 41 48.68 & +40 17 17.23 & 0.408 & 15.72 & 23.91 & 1.406 & 3.119 & 4 & \\[1pt]
PCG1044+0248 & 10 44 18.96 & +02 48 14.44 & 0.311 & 15.69 & 23.29 & 0.862 &   -   & 4 & \\[1pt]
PCG1044+3536 & 10 44 50.25 & +35 36  1.59 & 0.283 & 15.00 & 22.39 & 1.726 & 2.245 & 4 & \\[1pt]
PCG1045+4931 & 10 45 27.36 & +49 31 18.55 & 0.215 & 15.61 & 22.41 & 1.595 & 2.332 & 4 & \\[1pt]
PCG1045+2027 & 10 45 30.62 & +20 27  1.84 & 0.270 & 15.61 & 22.90 & 1.109 & 3.690 & 4 & \\[1pt]
PCG1045+1758 & 10 45 38.53 & +17 58 27.01 & 0.337 & 15.63 & 23.40 & 1.490 &   -   & 4 & \\[1pt]
PCG1054+1133 & 10 54  0.74 & +11 33 27.04 & 0.260 & 15.84 & 23.05 & 1.522 & 2.730 & 4 & \\[1pt]
PCG1100+0824 & 11 00  2.73 & +08 24 35.39 & 0.254 & 15.45 & 22.61 & 0.982 & 3.325 & 4 & \\[1pt]
PCG1120+0744 & 11 20 51.84 & +07 44 39.84 & 0.520 & 15.05 & 23.76 & 0.280 & 1.867 & 4 & \\[1pt]
PCG1123+3559 & 11 23 56.63 & +35 59 24.86 & 0.470 & 15.14 & 23.63 & 1.915 & 3.203 & 5 & \\[1pt]
PCG1137+3234 & 11 37  1.72 & +32 34 12.47 & 0.347 & 15.65 & 23.49 & 1.056 & 2.465 & 4 & \\[1pt]
PCG1151+2738 & 11 51 20.00 & +27 38  3.63 & 0.266 & 15.15 & 22.41 & 1.866 & 3.402 & 5 & \\[1pt]
PCG1156+0318 & 11 56 10.09 & +03 18  2.16 & 0.246 & 15.63 & 22.72 & 1.436 & 3.315 & 4 & \\[1pt]
PCG1212+2235 & 12 12 52.51 & +22 35 19.89 & 0.216 & 15.23 & 22.04 & 1.212 &   -   & 4 & \\[1pt]
PCG1221+5548 & 12 21 42.14 & +55 48 21.60 & 0.373 & 15.40 & 23.39 & 1.240 & 2.244 & 4 & \\[1pt]
PCG1222+1139 & 12 22 22.05 & +11 39 23.26 & 0.286 & 15.93 & 23.35 & 1.592 & 2.253 & 4 & \\[1pt]
PCG1352+1234 & 13 52 15.45 & +12 33 59.83 & 0.193 & 16.10 & 22.66 & 1.321 & 1.844 & 4 & \\[1pt]
PCG1513+1907 & 15 13 40.07 & +19 07 14.12 & 0.208 & 15.76 & 22.48 & 1.640 &   -   & 4 & \\[1pt]
PCG1516+0257 & 15 16 24.76 & +02 57 57.46 & 0.152 & 15.25 & 21.29 & 0.836 & 3.117 & 4 & \\[1pt]
PCG1525+2956 & 15 25  2.34 & +29 56  5.50 & 0.267 & 15.11 & 22.38 & 1.447 & 4.186 & 4 & \\[1pt]
PCG1528+4235 & 15 28 53.30 & +42 35 46.21 & 0.341 & 15.74 & 23.54 & 1.482 & 3.617 & 5 & \\[1pt]
PCG2221-0105 & 22 21 11.76 & -01 05 04.88 & 0.281 & 16.26 & 23.64 & 1.847 & 2.731 & 4 & \\[1pt]
PCG2226+0512 & 22 26 33.57 & +05 12 07.02 & 0.463 & 15.44 & 23.90 & 1.271 & 2.189 & 4 & \\[1pt]
PCG2259+1329 & 22 59  2.90 & +13 29 34.01 & 0.293 & 15.74 & 23.21 & 1.672 & 3.824 & 4 & \\[1pt]
PCG2312+1017 & 23 12 40.10 & +10 17 38.29 & 0.382 & 15.71 & 23.75 & 1.845 & 3.619 & 4 & \\[1pt]
PCG2324+0051 & 23 24 45.18 & +00 51 10.01 & 0.366 & 15.61 & 23.56 & 1.118 & 2.216 & 4 & \\[1pt]
PCG2328+0900 & 23 28  1.59 & +09 00 38.63 & 0.201 & 16.02 & 22.67 & 1.241 &   -   & 4 & \\[1pt]
PCG2332+1144 & 23 32 30.92 & +11 44 31.38 & 0.432 & 15.60 & 23.91 & 1.537 & 2.160 & 4 & \\[1pt]
PCG2334+0037 & 23 34 46.19 & +00 37 43.46 & 0.477 & 15.40 & 23.93 & 1.010 & 1.973 & 4 & \\[1pt]
PCG2350+1437 & 23 50 15.48 & +14 37 23.92 & 0.336 & 15.70 & 23.47 & 0.844 & 2.697 & 4 & \\[1pt]
\enddata
\end{deluxetable}

\begin{deluxetable}{cccccccccccc}
\tabletypesize{\scriptsize}
\tablecaption{Catalog of Groups}
\tablewidth{0pt}
\tablehead{
\colhead{Name} &\colhead{RA} &\colhead{DEC}& \colhead{m$_{\rm r}$} & \colhead{g-r}
& \colhead{PA}& \colhead{Ellip.} & \colhead{z}\\
\colhead{}&  \colhead{(J2000)}&  \colhead{(J2000)}     &  \colhead{mag}      & \colhead{mag}
&\colhead{$(\circ)$}&\colhead{}
}
\startdata
PCG0009+1958\\[1pt]
A & 00 09 55.14 & +19 58 27.66 & 16.861 & 0.341 &  38.0 & 0.206\\[1pt]
B & 00 09 53.83 & +19 58 11.21 & 17.421 & 0.174 &  61.4 & 0.023\\[1pt]
C & 00 09 55.21 & +19 58 45.44 & 18.174 & 0.384 & $-$70.6 & 0.053\\[1pt]
D & 00 09 55.36 & +19 58  5.88 & 18.456 & 0.382 &  63.3 & 0.146\\[1pt]
E & 00 09 53.15 & +19 58 37.85 & 18.695 & 0.316 & $-$71.2 & 0.205\\[1pt]
PCG0011+0544\\[1pt]
A & 00 11  9.08 & +05 44 44.20 & 16.811 & 0.309 &  86.3 & 0.203\\[1pt]
B & 00 11  9.80 & +05 44 54.06 & 17.729 & 0.514 & $-$18.9 & 0.047\\[1pt]
C & 00 11  8.14 & +05 44 44.16 & 17.972 & 0.019 &  81.1 & 0.276\\[1pt]
D & 00 11  8.17 & +05 44 51.40 & 18.630 & 0.180 &   2.2 & 0.177\\[1pt]
PCG0017$-$0206\\[1pt]
A & 00 17 11.09 & $-$02 06 29.52 & 16.323 & 0.261 &  33.6 & 0.007\\[1pt]
B & 00 17 13.96 & $-$02 05 48.88 & 16.405 & 0.277 & $-$75.6 & 0.123\\[1pt]
C & 00 17 13.62 & $-$02 06  6.19 & 16.669 & 0.088 &  43.0 & 0.188\\[1pt]
D & 00 17 13.07 & $-$02 06  2.09 & 17.483 & $-$0.104 &  $-$9.0 & 0.160\\[1pt]
PCG0038+0245\\[1pt]
A & 00 38 17.98 & +02 45 56.77 & 16.933 & 0.361 & $-$55.3 & 0.339\\[1pt]
B & 00 38 19.20 & +02 45 50.62 & 17.294 & 0.113 & $-$55.6 & 0.345\\[1pt]
C & 00 38 16.96 & +02 45 41.54 & 17.793 & 0.498 &  $-$8.7 & 0.047\\[1pt]
D & 00 38 18.93 & +02 45 21.06 & 18.606 & 0.113 &  71.7 & 0.440\\[1pt]
PCG0045+1940\\[1pt]
A & 00 45 19.61 & +19 39 59.94 & 16.547 & 0.483 &  34.6 & 0.237\\[1pt]
B & 00 45 18.61 & +19 39 57.92 & 17.669 & 0.787 & $-$45.3 & 0.499\\[1pt]
C & 00 45 20.04 & +19 40 14.02 & 17.676 & 0.392 & $-$51.1 & 0.124\\[1pt]
D & 00 45 19.61 & +19 40 13.01 & 17.945 & 0.424 &  20.6 & 0.164\\[1pt]
PCG0127+1459\\[1pt]
A & 01 27 35.14 & +14 59 25.48 & 16.610 & 0.299 &  57.4 & 0.119 & 0.110(sdss)\\[1pt]
B & 01 27 36.39 & +14 58 59.20 & 17.590 & 0.207 &  79.5 & 0.456 & \\[1pt]
C & 01 27 33.89 & +14 59 28.57 & 17.617 & 0.312 &  49.2 & 0.314 & 0.128(sdss)\\[1pt]
D & 01 27 35.84 & +14 59 32.46 & 18.168 & 0.360 &  63.8 & 0.292 & \\[1pt]
PCG0154+0139\\[1pt]
A & 01 54  4.15 & +01 39 32.33 & 16.192 & 0.406 &  21.0 & 0.255\\[1pt]
B & 01 54  2.13 & +01 40 14.99 & 16.209 & 0.440 &  45.0 & 0.479\\[1pt]
C & 01 54  2.78 & +01 39 24.48 & 16.380 & 0.450 &   2.2 & 0.124\\[1pt]
D & 01 53 59.95 & +01 39 36.79 & 16.696 & 0.484 & $-$33.2 & 0.306\\[1pt]
E & 01 54  3.20 & +01 39 23.94 & 16.763 & 0.422 & $-$66.1 & 0.274\\[1pt]
PCG0209+1039\\[1pt]
A & 02 09 14.83 & +10 39 46.30 & 16.350 & 0.484 &  $-$9.5 & 0.153\\[1pt]
B & 02 09 15.75 & +10 39 45.18 & 17.036 & 0.416 & $-$51.7 & 0.139\\[1pt]
C & 02 09 14.50 & +10 39 59.72 & 17.571 & 0.525 &  43.1 & 0.400\\[1pt]
D & 02 09 15.31 & +10 39 55.19 & 17.588 & 0.446 &  79.0 & 0.205\\[1pt]
PCG0209+0452\\[1pt]
A & 02 09 45.92 & +04 52 54.30 & 16.840 & 0.618 & $-$31.6 & 0.122\\[1pt]
B & 02 09 44.39 & +04 52 58.44 & 17.049 & 0.552 &  35.1 & 0.180\\[1pt]
C & 02 09 45.55 & +04 52 36.59 & 17.138 & 0.786 &  38.0 & 0.190\\[1pt]
D & 02 09 44.68 & +04 53  5.86 & 17.314 & 0.475 &  59.2 & 0.210\\[1pt]
PCG0250+0700\\[1pt]
A & 02 50 57.13 & +07 00 40.72 & 16.992 & 0.419 & $-$65.1 & 0.049\\[1pt]
B & 02 50 58.12 & +07 00 55.30 & 17.050 & 0.336 &  28.5 & 0.290\\[1pt]
C & 02 50 56.19 & +07 00 40.72 & 17.261 & 0.394 &  52.0 & 0.174\\[1pt]
D & 02 50 56.00 & +07 01  0.59 & 17.785 & 0.490 & $-$34.8 & 0.470\\[1pt]
E & 02 50 56.19 & +07 00 35.28 & 18.229 & 0.187 &  47.7 & 0.171\\[1pt]
PCG0303+0847\\[1pt]
A & 03 03 52.26 & +08 47  0.49 & 16.184 & 0.212 &  74.3 & 0.151\\[1pt]
B & 03 03 53.90 & +08 47 21.55 & 16.454 & 0.124 &  29.0 & 0.516\\[1pt]
C & 03 03 54.25 & +08 46 47.28 & 16.578 & 0.269 & $-$77.8 & 0.264\\[1pt]
D & 03 03 50.29 & +08 46 59.38 & 18.058 & $-$0.057 &  48.3 & 0.318\\[1pt]
PCG0854+4919\\[1pt]
A & 08 54 48.39 & +49 18 44.96 & 16.301 & 0.243 &  $-$6.9 & 0.172 & 0.118(sdss)\\[1pt]
B & 08 54 48.35 & +49 19 28.92 & 16.518 & 0.276 & $-$24.1 & 0.244\\[1pt]
C & 08 54 49.97 & +49 19 38.93 & 17.273 & 0.552 & $-$41.9 & 0.187 & 0.184(sdss)\\[1pt]
D & 08 54 47.41 & +49 19  2.56 & 17.420 & 0.298 &   0.6 & 0.146\\[1pt]
PCG0904+4523\\[1pt]
A & 09 04 26.92 & +45 23 38.40 & 16.917 & 0.562 & $-$77.6 & 0.021 & 0.137(sdss)\\[1pt]
B & 09 04 28.32 & +45 23 56.59 & 18.234 & 0.658 &  10.1 & 0.317\\[1pt]
C & 09 04 26.72 & +45 23 28.64 & 18.375 & 0.537 &  75.2 & 0.060\\[1pt]
D & 09 04 25.20 & +45 23 55.50 & 18.535 & 0.827 & $-$49.6 & 0.414\\[1pt]
PCG0915+2130\\[1pt]
A & 09 15 24.18 & +21 30 20.19 & 16.941 & 0.622 & $-$87.9 & 0.026\\[1pt]
B & 09 15 25.22 & +21 30 32.69 & 17.717 & 0.313 & $-$13.1 & 0.193\\[1pt]
C & 09 15 24.44 & +21 30 27.69 & 17.751 & 0.539 & $-$25.0 & 0.107\\[1pt]
D & 09 15 23.21 & +21 30 42.81 & 17.806 & 0.529 &  37.6 & 0.367\\[1pt]
E & 09 15 25.96 & +21 30 39.96 & 17.952 & 0.183 & $-$69.2 & 0.235\\[1pt]
PCG0922+2855\\[1pt]
A & 09 22 51.24 & +28 55 31.04 & 16.179 & 0.329 &  30.0 & 0.327 & 0.076\\[1pt]
B & 09 22 51.38 & +28 55 42.82 & 16.944 & 0.386 & $-$44.6 & 0.213\\[1pt]
C & 09 22 54.03 & +28 54 53.96 & 17.135 & 0.253 & $-$47.6 & 0.213\\[1pt]
D & 09 22 52.59 & +28 55  5.52 & 17.405 & 0.359 & $-$17.6 & 0.239\\[1pt]
PCG0928+6347\\[1pt]
A & 09 28 33.80 & +63 47 42.08 & 16.639 & 0.464 &  30.1 & 0.118\\[1pt]
B & 09 28 28.88 & +63 47 44.88 & 16.850 & 0.605 & $-$24.5 & 0.159\\[1pt]
C & 09 28 29.91 & +63 47 43.77 & 16.979 & 0.642 &  75.1 & 0.221\\[1pt]
D & 09 28 33.64 & +63 47 27.31 & 17.278 & 0.437 & $-$51.7 & 0.085\\[1pt]
E & 09 28 29.63 & +63 47 38.90 & 17.616 & 0.375 &  45.1 & 0.124\\[1pt]
PCG0939+1240\\[1pt]
A & 09 39 56.62 & +12 40 33.49 & 16.079 & 1.031 &  $-$7.4 & 0.088\\[1pt]
B & 09 39 58.14 & +12 40 31.22 & 16.610 & 1.199 & $-$79.5 & 0.385\\[1pt]
C & 09 39 55.21 & +12 40 44.22 & 16.638 & 1.193 & $-$50.8 & 0.310\\[1pt]
D & 09 39 54.21 & +12 40 44.22 & 16.910 & 1.059 &  18.0 & 0.125\\[1pt]
PCG0943+3923\\[1pt]
A & 09 43 18.02 & +39 22 53.00 & 16.980 & 0.393 & $-$50.3 & 0.104 & 0.151(sdss)\\[1pt]
B & 09 43 15.57 & +39 22 50.52 & 18.002 & $-$0.025 &  11.2 & 0.525\\[1pt]
C & 09 43 17.42 & +39 22 57.15 & 18.061 & 0.543 & $-$44.2 & 0.284\\[1pt]
D & 09 43 17.46 & +39 23 28.39 & 18.201 & 0.408 & $-$74.8 & 0.105\\[1pt]
E & 09 43 17.32 & +39 23 19.40 & 18.382 & 0.055 &  80.0 & 0.410\\[1pt]
PCG0953+5710\\[1pt]
A & 09 53 51.59 & +57 10 44.72 & 16.299 & 0.389 &  17.2 & 0.401 & 0.082(sdss)\\[1pt]
B & 09 53 55.22 & +57 10 43.72 & 16.541 & 0.353 & $-$20.2 & 0.222 & 0.081(sdss)\\[1pt]
C & 09 53 50.14 & +57 10 50.91 & 17.791 & 0.365 & $-$36.4 & 0.299\\[1pt]
D & 09 53 50.74 & +57 10 35.77 & 18.141 & 0.136 & $-$87.7 & 0.311\\[1pt]
PCG0955+0935\\[1pt]
A & 09 55  7.32 & +09 35 15.04 & 16.242 & 0.471 & $-$75.1 & 0.489\\[1pt]
B & 09 55  8.74 & +09 35 18.92 & 16.659 & 0.501 &  45.5 & 0.129\\[1pt]
C & 09 55  6.96 & +09 35  5.71 & 16.990 & 0.389 &  47.4 & 0.072\\[1pt]
D & 09 55  7.77 & +09 35 37.72 & 17.188 & 0.453 & $-$60.5 & 0.094\\[1pt]
PCG0955+0345\\[1pt]
A & 09 55 27.70 & +03 45 17.39 & 16.243 & 0.322 & $-$38.8 & 0.340 & 0.091(sdss)\\[1pt]
B & 09 55 26.29 & +03 45 26.10 & 16.594 & 0.050 & $-$80.7 & 0.239\\[1pt]
C & 09 55 28.14 & +03 44 50.57 & 16.971 & 0.313 &  12.2 & 0.232 & 0.094(sdss)\\[1pt]
D & 09 55 27.34 & +03 45 19.76 & 17.424 & 0.065 &   0.9 & 0.200\\[1pt]
PCG1003+1904\\[1pt]
A & 10 03 54.22 & +19 05  2.11 & 16.387 & 0.365 &  34.1 & 0.271\\[1pt]
B & 10 03 56.92 & +19 05  2.11 & 17.232 & 0.336 &  48.2 & 0.075\\[1pt]
C & 10 03 55.70 & +19 04 55.09 & 17.553 & 0.368 &  56.6 & 0.312\\[1pt]
D & 10 03 53.61 & +19 04 47.21 & 17.840 & 0.296 &  12.4 & 0.264\\[1pt]
PCG1008+1715\\[1pt]
A & 10 08 37.99 & +17 15 57.17 & 16.247 & 0.569 &  23.4 & 0.010\\[1pt]
B & 10 08 38.94 & +17 15 51.12 & 16.495 & 0.594 & $-$24.4 & 0.421 & 0.121\\[1pt]
C & 10 08 37.01 & +17 15 36.61 & 16.992 & 0.594 &  19.5 & 0.071\\[1pt]
D & 10 08 36.71 & +17 15 46.95 & 17.613 & 0.385 & $-$35.5 & 0.234\\[1pt]
PCG1010+0346\\[1pt]
A & 10 10 52.16 & +03 45 54.83 & 16.298 & 0.013 &  75.3 & 0.162 & 0.028(sdss)\\[1pt]
B & 10 10 55.15 & +03 46 31.01 & 16.684 & 0.303 &  55.2 & 0.056 & 0.095(sdss)\\[1pt]
C & 10 10 52.47 & +03 46  0.77 & 16.774 & 0.574 &  76.4 & 0.134\\[1pt]
D & 10 10 55.38 & +03 46 22.04 & 17.533 & 0.006 &  11.9 & 0.379 & 0.104(sdss)\\[1pt]
PCG1011+0841\\[1pt]
A & 10 11 14.92 & +08 41  8.12 & 16.208 & 0.286 & $-$18.0 & 0.313 & 0.097(sdss)\\[1pt]
B & 10 11 11.43 & +08 41 31.09 & 16.356 & 0.137 &  11.1 & 0.079\\[1pt]
C & 10 11 14.29 & +08 41 53.63 & 16.853 & 0.106 &  63.7 & 0.187\\[1pt]
D & 10 11 13.53 & +08 41 40.70 & 17.158 & $-$0.005 & $-$24.1 & 0.616\\[1pt]
PCG1013+2831\\[1pt]
A & 10 13 23.69 & +28 31 35.55 & 16.579 & 0.436 &  69.1 & 0.184\\[1pt]
B & 10 13 25.74 & +28 31 58.55 & 17.047 & 0.444 &  61.0 & 0.153\\[1pt]
C & 10 13 25.06 & +28 31 46.13 & 17.166 & 0.504 &  19.9 & 0.281\\[1pt]
D & 10 13 24.89 & +28 31 55.23 & 17.679 & 0.393 &  11.1 & 0.192\\[1pt]
PCG1013$-$0055\\[1pt]
A & 10 13 28.02 & $-$00 55 37.09 & 16.780 & 0.062 & $-$66.3 & 0.521\\[1pt]
B & 10 13 27.57 & $-$00 55 28.34 & 18.002 & 0.507 & $-$83.5 & 0.203\\[1pt]
C & 10 13 29.46 & $-$00 55 27.37 & 18.148 & 0.324 &  88.8 & 0.430\\[1pt]
D & 10 13 27.96 & $-$00 55 26.54 & 18.187 & 0.795 & $-$66.7 & 0.254\\[1pt]
E & 10 13 29.89 & $-$00 55 15.71 & 18.381 & $-$0.139 &  35.2 & 0.385\\[1pt]
PCG1021+3713\\[1pt]
A & 10 21 57.03 & +37 13 16.54 & 16.723 & 0.546 &  86.6 & 0.322\\[1pt]
B & 10 21 57.42 & +37 13  3.90 & 17.393 & 0.382 & $-$42.6 & 0.102\\[1pt]
C & 10 21 57.29 & +37 13  8.26 & 17.422 & 0.426 &  82.0 & 0.309\\[1pt]
D & 10 21 57.23 & +37 13 37.12 & 17.846 & 0.426 &  78.0 & 0.292\\[1pt]
PCG1025+3601\\[1pt]
A & 10 25 44.47 & +36 01 21.86 & 16.317 & 0.193 &   2.2 & 0.138\\[1pt]
B & 10 25 44.23 & +36 01 39.98 & 17.061 & 0.923 &  84.1 & 0.107\\[1pt]
C & 10 25 43.95 & +36 01 23.81 & 17.826 & $-$0.204 & $-$43.6 & 0.345\\[1pt]
D & 10 25 45.14 & +36 01 45.80 & 18.164 & 0.990 &  37.5 & 0.170\\[1pt]
PCG1026+3906\\[1pt]
A & 10 26  7.91 & +39 06  5.03 & 16.708 & 0.421 &   6.2 & 0.235\\[1pt]
B & 10 26  7.61 & +39 06 14.29 & 17.389 & 0.458 &  87.8 & 0.094\\[1pt]
C & 10 26  8.33 & +39 06  1.88 & 17.881 & 0.601 &  77.3 & 0.292\\[1pt]
D & 10 26  7.91 & +39 05 58.46 & 18.342 & 0.232 &  35.8 & 0.096\\[1pt]
PCG1041+4017\\[1pt]
A & 10 41 48.02 & +40 17 21.08 & 16.734 & 0.253 &  32.7 & 0.168 & 0.069\\[1pt]
B & 10 41 49.64 & +40 16 55.30 & 16.788 & 0.091 & $-$16.1 & 0.455\\[1pt]
C & 10 41 46.95 & +40 17 31.64 & 17.972 & 0.092 &  40.0 & 0.525\\[1pt]
D & 10 41 49.92 & +40 17 37.24 & 18.140 & 0.523 &  32.4 & 0.247\\[1pt]
PCG1044+0248\\[1pt]
A & 10 44 17.91 & +02 48  4.32 & 16.870 & $-$0.115 & $-$72.7 & 0.290\\[1pt]
B & 10 44 20.17 & +02 48 18.97 & 16.887 & $-$0.146 & $-$68.0 & 0.163\\[1pt]
C & 10 44 18.08 & +02 48 18.04 & 17.542 & 0.498 &  38.4 & 0.131 & 0.123(sdss)\\[1pt]
D & 10 44 20.01 & +02 48 24.59 & 17.732 & 0.720 &  $-$8.7 & 0.202\\[1pt]
PCG1044+3536\\[1pt]
A & 10 44 50.95 & +35 36  6.30 & 16.021 & $-$0.071 &  45.6 & 0.350 & 0.051\\[1pt]
B & 10 44 50.05 & +35 36 18.40 & 16.052 & 0.332 &  $-$3.4 & 0.534\\[1pt]
C & 10 44 51.40 & +35 35 51.97 & 17.066 & 0.529 & $-$58.8 & 0.108\\[1pt]
D & 10 44 48.87 & +35 35 59.21 & 17.747 & 0.313 & $-$62.9 & 0.224\\[1pt]
PCG1045+4931\\[1pt]
A & 10 45 26.76 & +49 31  7.03 & 16.446 & 0.565 & $-$83.8 & 0.107 & 0.173(sdss)\\[1pt]
B & 10 45 26.25 & +49 31 25.68 & 17.005 & 0.511 &  77.9 & 0.191\\[1pt]
C & 10 45 28.12 & +49 31 12.76 & 17.632 & 0.515 &  43.9 & 0.208 & 0.168(sdss)\\[1pt]
D & 10 45 28.65 & +49 31 15.60 & 18.041 & 0.531 & $-$29.7 & 0.594\\[1pt]
PCG1045+2027\\[1pt]
A & 10 45 30.79 & +20 27 17.82 & 16.520 & 0.326 &  65.7 & 0.512\\[1pt]
B & 10 45 30.67 & +20 26 53.09 & 17.267 & 0.276 &  $-$0.5 & 0.074\\[1pt]
C & 10 45 31.19 & +20 26 55.07 & 17.418 & 0.173 &  44.4 & 0.152\\[1pt]
D & 10 45 30.45 & +20 26 45.82 & 17.629 & 0.276 &  36.3 & 0.162\\[1pt]
PCG1045+1758\\[1pt]
A & 10 45 39.91 & +17 58 31.48 & 16.647 & 0.231 &   2.5 & 0.340\\[1pt]
B & 10 45 38.98 & +17 58  7.86 & 16.651 & 0.290 & $-$32.6 & 0.245\\[1pt]
C & 10 45 37.78 & +17 58 17.55 & 17.940 & 0.255 & $-$37.9 & 0.333\\[1pt]
D & 10 45 37.90 & +17 58 45.16 & 18.137 & 0.312 &  48.6 & 0.302\\[1pt]
PCG1054+1133\\[1pt]
A & 10 53 59.93 & +11 33 16.89 & 16.697 & 0.072 &  30.2 & 0.088\\[1pt]
B & 10 54  1.79 & +11 33 26.28 & 17.139 & 0.246 &  60.2 & 0.107\\[1pt]
C & 10 54  0.69 & +11 33 36.76 & 18.022 & 0.464 &  40.8 & 0.379\\[1pt]
D & 10 54  1.15 & +11 33 41.40 & 18.219 & 0.102 &  38.0 & 0.154\\[1pt]
PCG1100+0824\\[1pt]
A & 11 00  1.85 & +08 24 27.47 & 16.493 & 0.065 &  70.0 & 0.318\\[1pt]
B & 11 00  2.03 & +08 24 44.71 & 17.040 & 0.410 &  68.8 & 0.190\\[1pt]
C & 11 00  3.61 & +08 24 43.31 & 17.058 & 0.263 &  23.3 & 0.215 & 0.073(sdss)\\[1pt]
D & 11 00  3.11 & +08 24 48.28 & 17.475 & 0.297 &  57.0 & 0.067\\[1pt]
PCG1120+0744\\[1pt]
A & 11 20 49.96 & +07 44 25.98 & 16.446 & 0.184 & $-$86.1 & 0.634\\[1pt]
B & 11 20 50.48 & +07 45  3.67 & 16.458 & 0.407 &   0.5 & 0.293\\[1pt]
C & 11 20 53.88 & +07 44 32.78 & 16.583 & 0.437 & $-$59.6 & 0.296\\[1pt]
D & 11 20 51.38 & +07 45  2.30 & 16.726 & 0.286 &   7.4 & 0.652\\[1pt]
PCG1123+3559\\[1pt]
A & 11 23 55.11 & +35 59 28.14 & 16.069 & 0.085 &  24.4 & 0.662\\[1pt]
B & 11 23 55.17 & +35 59 13.88 & 16.728 & 0.075 &  43.8 & 0.394\\[1pt]
C & 11 23 54.45 & +35 59 15.00 & 17.218 & 0.342 &  $-$7.6 & 0.282\\[1pt]
D & 11 23 58.86 & +35 59 24.36 & 17.375 & 0.308 &  68.2 & 0.060\\[1pt]
E & 11 23 58.80 & +35 59 34.73 & 17.984 & 0.297 & $-$13.8 & 0.486\\[1pt]
PCG1137+3234\\[1pt]
A & 11 37  0.94 & +32 34  6.45 & 16.778 & 0.617 &  12.9 & 0.190\\[1pt]
B & 11 37  0.11 & +32 34 16.90 & 16.944 & 0.495 &  83.5 & 0.318\\[1pt]
C & 11 37  0.48 & +32 33 58.72 & 17.389 & 0.740 &  88.3 & 0.250\\[1pt]
D & 11 37  3.35 & +32 34  9.55 & 17.834 & 0.491 & $-$38.2 & 0.453\\[1pt]
PCG1151+2738\\[1pt]
A & 11 51 21.21 & +27 38  3.80 & 16.214 & 0.015 &  47.3 & 0.611\\[1pt]
B & 11 51 20.07 & +27 38 19.57 & 16.585 & 0.211 & $-$81.2 & 0.118\\[1pt]
C & 11 51 19.25 & +27 37 51.17 & 17.144 & 0.313 & $-$13.7 & 0.042\\[1pt]
D & 11 51 19.69 & +27 38 15.72 & 17.372 & 0.397 &  61.5 & 0.218\\[1pt]
E & 11 51 19.54 & +27 38 10.64 & 18.080 & 0.113 &  70.2 & 0.365\\[1pt]
PCG1156+0318\\[1pt]
A & 11 56  9.76 & +03 17 48.70 & 16.298 & 0.111 &  39.3 & 0.288\\[1pt]
B & 11 56  9.11 & +03 18  4.97 & 17.577 & 0.249 &  87.6 & 0.294\\[1pt]
C & 11 56 10.50 & +03 17 57.73 & 17.665 & 0.231 &  16.4 & 0.394 & 0.072(sdss)\\[1pt]
D & 11 56 11.05 & +03 17 59.35 & 17.734 & 0.173 & $-$54.9 & 0.133\\[1pt]
PCG1212+2235\\[1pt]
A & 12 12 53.31 & +22 35 26.55 & 16.393 & 0.147 &  48.5 & 0.409\\[1pt]
B & 12 12 51.93 & +22 35 30.08 & 16.460 & 0.056 &  35.3 & 0.116\\[1pt]
C & 12 12 52.42 & +22 35 14.28 & 16.877 & 0.435 & $-$59.1 & 0.161\\[1pt]
D & 12 12 52.55 & +22 35  6.94 & 17.605 & 0.396 & $-$88.8 & 0.232\\[1pt]
PCG1221+5548\\[1pt]
A & 12 21 39.72 & +55 48 12.38 & 16.236 & 0.180 &  30.2 & 0.727 & 0.044\\[1pt]
B & 12 21 43.67 & +55 48 39.85 & 17.131 & 0.193 & $-$49.8 & 0.643\\[1pt]
C & 12 21 42.56 & +55 48 32.37 & 17.227 & 0.363 &  66.6 & 0.099 & 0.034(sdss)\\[1pt]
D & 12 21 44.51 & +55 48 11.52 & 17.476 & 0.316 &   5.4 & 0.132\\[1pt]
PCG1222+1139\\[1pt]
A & 12 22 22.60 & +11 39 38.38 & 16.860 & 0.461 & $-$47.7 & 0.066\\[1pt]
B & 12 22 23.18 & +11 39 19.33 & 17.427 & 0.613 & $-$60.9 & 0.245\\[1pt]
C & 12 22 22.49 & +11 39 14.26 & 17.536 & 0.494 &  87.5 & 0.150\\[1pt]
D & 12 22 21.19 & +11 39 11.52 & 18.452 & 0.245 &  34.5 & 0.271\\[1pt]
PCG1352+1233\\[1pt]
A & 13 52 14.66 & +12 34  0.19 & 16.978 & 0.388 &  24.4 & 0.076\\[1pt]
B & 13 52 15.28 & +12 33 55.19 & 17.633 & 0.391 &  $-$7.3 & 0.082\\[1pt]
C & 13 52 15.77 & +12 33 59.18 & 17.935 & 0.392 &  $-$9.1 & 0.273\\[1pt]
D & 13 52 16.24 & +12 33 59.47 & 18.299 & 0.314 & $-$57.0 & 0.144\\[1pt]
PCG1513+1907\\[1pt]
A & 15 13 40.26 & +19 07 21.65 & 16.734 & $-$0.050 & $-$14.6 & 0.080\\[1pt]
B & 15 13 39.61 & +19 07 11.67 & 17.187 & 0.238 &   2.5 & 0.134\\[1pt]
C & 15 13 39.76 & +19 07 25.75 & 17.332 & 0.461 &  65.5 & 0.031\\[1pt]
D & 15 13 40.39 & +19 07  2.49 & 18.374 & 0.297 & $-$10.4 & 0.351\\[1pt]
PCG1516+0257\\[1pt]
A & 15 16 25.25 & +02 57 56.59 & 16.512 & 0.623 & $-$80.8 & 0.314\\[1pt]
B & 15 16 25.31 & +02 58  1.31 & 16.548 & 0.633 & $-$87.4 & 0.208 & 0.113(sdss)\\[1pt]
C & 15 16 24.20 & +02 57 53.60 & 16.768 & 0.593 & $-$42.7 & 0.139\\[1pt]
D & 15 16 24.75 & +02 58  1.60 & 17.348 & 0.492 &  20.2 & 0.337\\[1pt]
PCG1525+2956\\[1pt]
A & 15 25  2.97 & +29 55 51.78 & 16.221 & 0.314 &  $-$6.5 & 0.225\\[1pt]
B & 15 25  2.04 & +29 55 56.96 & 16.388 & 0.395 &  69.6 & 0.072\\[1pt]
C & 15 25  2.21 & +29 56 21.41 & 16.670 & 0.541 &  25.7 & 0.230\\[1pt]
D & 15 25  1.41 & +29 55 54.98 & 17.668 & 0.622 &  12.3 & 0.050\\[1pt]
PCG1528+4235\\[1pt]
A & 15 28 51.54 & +42 35 40.05 & 16.717 & 0.372 & $-$16.7 & 0.130\\[1pt]
B & 15 28 52.36 & +42 35 58.17 & 17.559 & 0.286 &  27.1 & 0.273\\[1pt]
C & 15 28 54.19 & +42 35 40.96 & 17.748 & 0.208 &  39.5 & 0.186\\[1pt]
D & 15 28 54.58 & +42 36  1.04 & 17.872 & 0.304 &  20.5 & 0.465\\[1pt]
E & 15 28 54.20 & +42 35 28.35 & 18.199 & 0.273 & $-$63.0 & 0.555\\[1pt]
PCG2221$-$0105\\[1pt]
A & 22 21 12.42 & $-$01 05  6.40 & 16.950 & 0.358 & $-$23.4 & 0.567 & 0.107(sdss)\\[1pt]
B & 22 21 10.68 & $-$01 05  9.46 & 17.912 & 0.819 & $-$86.5 & 0.279\\[1pt]
C & 22 21 10.74 & $-$01 04 57.36 & 18.274 & 0.794 & $-$26.9 & 0.214\\[1pt]
D & 22 21 12.76 & $-$01 05 12.44 & 18.797 & 0.418 &  57.5 & 0.173\\[1pt]
PCG2226+0512\\[1pt]
A & 22 26 34.35 & +05 11 47.40 & 16.346 & 0.206 & $-$56.3 & 0.183\\[1pt]
B & 22 26 35.43 & +05 12  8.57 & 16.785 & 0.332 & $-$52.4 & 0.211\\[1pt]
C & 22 26 31.71 & +05 12  5.47 & 17.579 & $-$0.053 &  60.6 & 0.165\\[1pt]
D & 22 26 32.72 & +05 12 11.56 & 17.617 & 0.281 & $-$14.4 & 0.503\\[1pt]
PCG2259+1329\\[1pt]
A & 22 59  3.95 & +13 29 25.40 & 16.467 & 0.467 &  55.9 & 0.187 & 0.129(sdss)\\[1pt]
B & 22 59  2.16 & +13 29 37.46 & 17.408 & 0.441 & $-$59.4 & 0.357\\[1pt]
C & 22 59  1.85 & +13 29 42.61 & 17.736 & 0.251 &  83.1 & 0.197\\[1pt]
D & 22 59  2.66 & +13 29 33.90 & 18.139 & 0.268 & $-$37.8 & 0.292\\[1pt]
PCG2312+1017\\[1pt]
A & 23 12 41.65 & +10 17 40.77 & 16.668 & 0.333 & $-$61.4 & 0.249\\[1pt]
B & 23 12 39.79 & +10 18  0.76 & 17.104 & 0.331 & $-$13.8 & 0.517\\[1pt]
C & 23 12 40.88 & +10 17 26.63 & 17.260 & 0.271 &  13.7 & 0.143\\[1pt]
D & 23 12 38.56 & +10 17 35.34 & 18.513 & $-$0.096 & $-$50.8 & 0.309\\[1pt]
PCG2324+0051\\[1pt]
A & 23 24 46.17 & +00 50 54.46 & 16.636 & 0.423 &  50.0 & 0.114\\[1pt]
B & 23 24 45.83 & +00 50 50.35 & 17.131 & 0.369 & $-$56.8 & 0.426\\[1pt]
C & 23 24 43.73 & +00 51 11.70 & 17.242 & 0.231 & $-$19.2 & 0.438\\[1pt]
D & 23 24 46.65 & +00 51  9.76 & 17.754 & 0.475 & $-$18.7 & 0.516 & 0.119(sdss)\\[1pt]
PCG2328+0900\\[1pt]
A & 23 28  1.94 & +09 00 44.42 & 16.912 & 0.431 &  36.0 & 0.068\\[1pt]
B & 23 28  1.99 & +09 00 36.97 & 17.605 & 0.489 & $-$25.0 & 0.164\\[1pt]
C & 23 28  1.59 & +09 00 50.72 & 17.835 & 0.390 &  $-$9.7 & 0.030\\[1pt]
D & 23 28  1.58 & +09 00 26.57 & 18.153 & 0.364 & $-$77.6 & 0.265\\[1pt]
PCG2332+1144\\[1pt]
A & 23 32 32.66 & +11 44 35.70 & 16.448 & 0.368 & $-$88.5 & 0.050\\[1pt]
B & 23 32 29.18 & +11 44 27.05 & 17.245 & 0.486 & $-$66.2 & 0.169\\[1pt]
C & 23 32 31.06 & +11 44 20.69 & 17.278 & 0.043 &  78.8 & 0.624\\[1pt]
D & 23 32 31.06 & +11 44 35.99 & 17.985 & 0.389 &  49.2 & 0.388\\[1pt]
PCG2334+0037\\[1pt]
A & 23 34 45.71 & +00 38  0.46 & 16.655 & 0.351 & $-$66.6 & 0.183\\[1pt]
B & 23 34 45.37 & +00 37 17.65 & 16.778 & 0.274 & $-$30.0 & 0.112\\[1pt]
C & 23 34 44.49 & +00 37 56.39 & 16.817 & 0.351 &  87.8 & 0.259 & 0.086(sdss)\\[1pt]
D & 23 34 47.95 & +00 37 54.30 & 17.665 & 0.181 & $-$87.2 & 0.350\\[1pt]
PCG2350+1437\\[1pt]
A & 23 50 14.27 & +14 37 14.20 & 16.665 & 0.654 &  $-$2.4 & 0.055\\[1pt]
B & 23 50 15.82 & +14 37 43.47 & 17.421 & 0.647 & $-$87.6 & 0.075\\[1pt]
C & 23 50 14.97 & +14 37 33.46 & 17.501 & 0.935 &   2.6 & 0.191\\[1pt]
D & 23 50 16.86 & +14 37 26.65 & 17.509 & 0.182 &  17.4 & 0.254 & 0.201(sdss)
\enddata
\end{deluxetable}
\end{document}